\documentclass[12pt]{article}
\usepackage{rotating} 
\usepackage{adjustbox,lipsum}
\usepackage{float}
\usepackage{amssymb}
\usepackage{amsthm}
\usepackage{amsmath}
\usepackage[noend]{algpseudocode}
\usepackage{latexsym}
\usepackage{epsfig}
\usepackage{graphicx}
\usepackage{wasysym}
\usepackage{threeparttable}
\usepackage{natbib}
\usepackage{color}
\usepackage{epstopdf}
\usepackage{caption}
\usepackage{subcaption}
\usepackage{booktabs}

\usepackage{multirow}
\usepackage[breaklinks]{hyperref}

\usepackage{enumitem}
\usepackage{caption}
\usepackage{algorithm}

\def\boxit#1{\vbox{\hrule\hbox{\vrule\kern6pt
          \vbox{\kern6pt#1\kern6pt}\kern6pt\vrule}\hrule}}

\def\bse{\begin{eqnarray*}}
\def\ese{\end{eqnarray*}}
\def\be{\begin{eqnarray}}
\def\ee{\end{eqnarray}}
\def\bq{\begin{equation}}
\def\eq{\end{equation}}
\def\bse{\begin{eqnarray*}}
\def\ese{\end{eqnarray*}}

\usepackage{mathptmx}      
\usepackage{bm}

 {\begin{list}{}%
         {\setlength{\leftmargin}{#1}}%
         \item[]%
 }
 {\end{list}}
\usepackage{setspace}

\begin{document}
\thispagestyle{empty} \baselineskip=28pt

\begin{center}
{\LARGE{\bf What is the best predictor that you can compute in five minutes using a given Bayesian hierarchical model?}}
\end{center}

\baselineskip=12pt

\vskip 2mm
\begin{center}
Jonathan R. Bradley\footnote{(\baselineskip=10pt to whom correspondence should be addressed) Department of Statistics, Florida State University, 117 N. Woodward Ave., Tallahassee, FL 32306-4330, jrbradley@fsu.edu}
\end{center}
%
%
%
%
\vskip 4mm

\begin{center}
\large{{\bf Abstract}}
\end{center}
The goal of this paper is to provide a way for statisticians to answer the question posed in the title of this article using any Bayesian hierarchical model of their choosing and without imposing additional restrictive model assumptions. We are motivated by the fact that the rise of ``big data'' has created difficulties for statisticians to directly apply their methods to big datasets. We introduce a ``data subset model'' to the popular ``data model, process model, and parameter model'' framework used to summarize Bayesian hierarchical models. The hyperparameters of the data subset model are specified constructively in that they are chosen such that the implied size of the subset satisfies pre-defined computational constraints. Thus, these hyperparameters effectively calibrates the statistical model to the computer itself to obtain predictions/estimations in a pre-specified amount of time. Several properties of the data subset model are provided including: propriety, partial sufficiency, and semi-parametric properties. Furthermore, we show that subsets of normally distributed data are asymptotically partially sufficient under reasonable constraints. Results from a simulated dataset will be presented across different computers, to show the effect of the computer on the statistical analysis. Additionally, we provide a joint spatial analysis of two different environmental datasets.
\baselineskip=12pt

%
%
%

\baselineskip=12pt
\par\vfill\noindent
\par\medskip\noindent
\clearpage\pagebreak\newpage \pagenumbering{arabic}
\baselineskip=24pt
\section{Introduction}
%
%

The computational difficulties involved with ``big data'' can often be reduced to difficulties involved with searching through a big parameter space. Consequently, there are an exorbitant number of methods for big data that reduce the parameter space to something that is easier to search through. 
For example, dimension reduction is common in principal component analysis, spatial analysis, and spatio-temporal analysis \citep[e.g, see][among several others]{jolliffe1973discarding,wiklecress_spt,johan-2006, cressie-tao,banerjee, johan,finley,katzfuss_1, cressie-shi-kang-2010,kang-cressie-shi-2010,kang-cressie-2011,katzfuss2012,bradleyMSTM}, which involves setting ``small'' eigenvalues of a covariance matrix equal to zero (reducing the parameter space). Another example is covariance tapering, which involves removing parameters by enforcing sparsity among the elements of a covariance matrix \citep[e.g., see][among several others]{furrer2006covariance,friedman2008sparse,rothman2010sparse,bien2011sparse}. These two example strategies place the burden of computation on the statistical model, and \textit{not} on the computer itself. 

An alternative approach is to subsample the data \citep{broderick2013streaming,srivastava2018scalable,zhao2018efficient} so that the parameter space associated with the subsample has a smaller (than the full data) parameter space (i.e., if big data has a big parameter space then small data has a small parameter space). This approach does not place the computational burden on the model/parameter space. In this article, we develop a data subsampling method that has several desirable properties. Specifically, our proposed approach (1) has a hyperparameter that can be selected to achieve a computational goal (e.g., inference in five minutes), (2) can be reasonably developed for any Bayesian hierarchical model (i.e., does not require additional assumptions), (3) can be applied to a dataset of arguably \textit{any} size, and (4) is an exact method. We argue that these properties are needed to shift the burden of computation from the statistical model to the computer itself.

The primary goal of this article is to answer the question posed in this article \textit{for any} given Bayesian hierarchical model, which we write using the general ``data, process, and parameter model'' notation \citep[e.g., see][among others]{berlinhier,cressie-wikle-book}. Specifically, we introduce additional levels to a generic Bayesian hierarchical model that leads to a subsampling procedure in its implementation. These new levels introduced to the hierarchical model are called the ``data subset model'' and the ``subset model.'' The data subset model introduces Bernoulli random variables that either keeps a datum's contribution to the likelihood, or removes it. The likelihood is reweighted in a manner that leads to a proper model, enforces the same assumptions on the latent processes, parameters, and data as the traditional hierarchical model (i.e., without subsampling). Our model implies that a subsample of the dataset is a partially sufficient statistic \citep[see][for a review]{BasuReview}. This allows for a straightforward Markov chain Monte Carlo (MCMC) method to generate values from the posterior distribution \citep{gelfand1990sampling} that bares similarity to existing methods in the machine learning literature \citep[e.g., see][among others]{kleiner2012big,Bardenet,korattikara2014austerity}.

The literature is teeming with existing approaches for subsampling data. One popular approach is referred to as ``divide and conquer'' \citep{kleiner2012big,LIANGsub,BARBIAN}. These methods subsample the data, then the model is fitted to each subsample, and the results are combined across the subsets. An alternative approach is to partition the data by regions (e.g., for spatial datasets), and assume independence across regions (but not within). There are several options to partition a spatial domain. For example, there are methods based on regular partitions \citep{sang}, hierarchical clustering of gradients \citep{ANDERSONsub, HEATONsub}, and there are methods based on partitions defined by the clustering of centroids of areal units \citep{KNORR,kim2}. There are also mixture models \citep{NEELON} and tree-based regression methods \citep{KONOMI} available. A clear limitation of these approaches, however, is that the assumption of blockwise independence can be extremely strong \citep[e.g., see discussion in][among others]{bradleyTEST}. 

Several methods reduce the parameter space in a way that effectively subsets the data in its implementation. For example, assuming conditional independence has been used to effectively subset the data. Gibbs distributions \citep[e.g., ][Chapter 4]{banerjee-etal-2004} assume conditional independence between an observation and observations outside a neighborhood, given observations within a neighborhood. Such models are referred to as Markov random fields \citep[e.g., see][for a standard reference]{Rue2005gaussian}. The conditional autoregressive (CAR) model developed by \citet{besag-74} is a specific type of Markov random field. \citet{Geman1984stochastic}, \citet{besag-86}, and \citet{besag-91} has made the CAR model a standard approach for image restoration.


The choice to use a convenient parsimonious model can be problematic. This is because ``convenient parsimony'' can be confused with ``realistic convenient parsimony.'' For example, reduced rank methods have been a popular method that simplifies a model, but has been shown to lead to inferential issues \citep{steinr,bradley2011}. Specifically, incorporating too few spatial random effects can speed up estimation and prediction \citep{johan,banerjee}, but can lead to problems with over-smoothing and uncertainty quantification \citep[e.g., see][for an example]{heaton2018case}. These difficulties with reduced rank models have only been shown when the data are Gaussian, spatially-referenced, no covariates are available, and the reduced rank expansion is used in the first level of the hierarchical model \citep{steinr}. Consequently, it is not entirely clear when a reduced rank model is appropriate, which creates more ambiguity. Our approach avoids possible model miss-specification in the name of parsimony.

Ideally, we would like to simultaneously model all the data without imposing additional assumptions on the dependence structure/parameter space. This would allow one to consider parametric structures that are not computationally convenient. There is a method available that does this, which is referred to as ``firefly MCMC'' \citep{Firefly2014}. This approach involves finding an easy to compute density that approximates the likelihood. However, in practice, it is not always clear how one specifies this easy to compute density. Also, several approximate methods exist as well, which  estimate a generic $N$-dimensional likelihood using $n<N$ data values \citep[e.g.,][among several others]{vecchia1988estimation,stein2004approximating,huggins2016coresets,gunawan2017fast,katzfuss2017general,srivastava2018scalable,guinness2018permutation,quiroz2019speeding,dang2019hamiltonian}.

Our method is exact, and is specified in a flexible semi-parametric manner. Specifically, when subsetting the data we assume a parametric model for the marginal distribution of the subset, and place very few assumptions on the data generating mechanism for the marginal distribution of the observations left out of the subset. Thus, we have a new semi-parametric Bayesian model. We make a distinction with the typical semi-parametric Bayesian literature. In particular, semi-parametric Bayesian methods are often framed as infinite-dimensional (or infinitely parametric) such that the implied parameter space is so large that it contains the true data generating mechanism \citep{sethuraman1994constructive}. For our setting, by ``semi-parametric'' we mean that the (marginal) data generating mechanism of the holdout data is left unknown and unparameterized (hence semi-parametric). 

The remainder of this article is organized as follows. In Section 2, we introduce the data subset model and a subset model into the Bayesian hierarchical model framework. We also develop several properties, provide computational details, and give example models within our framework. Then, in Section~3 we provide a simulation study to show the performance of the data subset model across multiple computers. Additionally, in Section~4 we provide a joint spatial analysis of two different high-dimensional environmental datasets. Finally, Section~5 contains a discussion. All proofs are given in the Appendix.

\section{Methodology}
Suppose we observe an $N$-dimensional data vector $\textbf{y} = (Y_{1},\ldots, Y_{N})^{\prime}$. We organize the latent process of interest into an $N$-dimensional vector $\bm{\nu}$. The joint distribution of the data, process, and parameters can succinctly be summarized using the ``data model, process, and parameter model'' terminology often used in the spatio-temporal statistics literature \citep[e.g., see][]{berlinhier,cressie-wikle-book}. In particular, the ``data model'' refers to the conditional distribution $f(\textbf{y}\vert \bm{\nu},\bm{\theta})$, where $f$ will be used to denote a probability density function/probability mass function (pdf/pmf) and $\bm{\theta}$ is a generic real-valued parameter vector. A hierarchical model can be written as the product of the following conditional and marginal distributions:
 	\begin{align}
 	\nonumber
 	&\mathrm{Data\hspace{5pt}Model:}\hspace{5pt} \prod_{i = 1}^{N} f(Y_{i} \vert \bm{\nu},\bm{\theta})\\
 	\nonumber
 	&\mathrm{Process\hspace{5pt}Model:}\hspace{5pt} f(\bm{\nu}\vert \bm{\theta})\\
 	\label{hmnotation}
 	&\mathrm{Parameter\hspace{5pt}Model:} \hspace{5pt}f(\bm{\theta}).
 	\end{align}
 \noindent
 We consider the setting where $N$ is so large that estimating $\bm{\nu}$ and $\bm{\theta}$ from (\ref{hmnotation}) directly is not possible. We henceforth refer to (\ref{hmnotation}) as the ``full parametric model,'' since $Y_{1},\ldots, Y_{N}$ all contribute to the likelihood implied by (\ref{hmnotation}).

\subsection{Bayesian Hierarchical Models with a data subset model: The Reduced Model}Our approach to data subsampling involves defining a mixture distribution over all possible subsets of the partitions. Consider the following fully Bayesian hierarchical model:
\begin{align}
\nonumber
&\mathrm{Data \hspace{5pt}Subset\hspace{5pt}Model:}\hspace{5pt} \left\lbrace\prod_{i = 1}^{N} f(Y_{i}\vert \bm{\nu},\bm{\theta},\bm{\delta})^{\delta_{i}} \right\rbrace\hspace{5pt} \frac{m\left(\bm{1}_{N},\textbf{y}\right)}{m(\bm{\delta}, \textbf{y}_{\delta})}\\
\nonumber
&\mathrm{Process\hspace{5pt}Model:}\hspace{5pt} f(\bm{\nu}\vert \bm{\theta})\\
\nonumber
&\mathrm{Parameter\hspace{5pt}Model:} \hspace{5pt}f(\bm{\theta})\\
\label{subhmnotation}
&\mathrm{Subset\hspace{5pt}Model:}\hspace{5pt}Pr(\bm{\delta}\vert n),
\end{align}
\noindent
where the 
$N$-dimensional random vector $\bm{\delta}=(\delta_{i},\ldots,\delta_{N})^{\prime}$ consists of ones and zeros, $\bm{1}_{N}$ (and $\bm{0}_{N}$) is an $N$-dimensional vector of ones (zeros), and the $n$-dimensional random vector $\textbf{y}_{\delta}= \left(Y_{i}: \delta_{i}=1\right)^{\prime}$. Set $\sum_{i = 1}^{N} \delta_{i} = n$, and we choose the value of $n$ in a manner that achieves the computational constraints (see Sections 2.3 and 2.4). We specify $f(Y_{i}\vert \bm{\nu},\bm{\theta},\bm{\delta} = \bm{1}_{N})$ in (\ref{subhmnotation}) to equal $f(Y_{i}\vert \bm{\nu},\bm{\theta})$ in (\ref{hmnotation}). When $\delta_{i} = 0$ the component of the joint distribution that contains $Y_{i}$ does not contain $\bm{\nu}$ and $\bm{\theta}$ in its expression, and vice versa. Consequently, $\delta_{i} = 0$ removes $Y_{i}$ from the expression of the likelihood (proportionally), and $\delta_{i} = 1$ includes $Y_{i}$ in the expression of the likelihood. As a result, we henceforth refer to (\ref{subhmnotation}) as the ``reduced parametric model.'' There are several choices for $Pr(\bm{\delta}\vert n)$ that one can take from the survey sampling literature \citep{lohr-survey}. In this article, we consider specifying $Pr(\bm{\delta}\vert n)$ to produce a simple random sample without replacement. However, complex survey designs could be used in settings not considered in this article. 

The data subset model contains two additional functions not included in the full parametric model. Specifically, define the marginal distribution:
\begin{equation}
m(\bm{\delta},\textbf{y})=\int \int \left\lbrace\prod_{i = 1}^{N} f(Y_{i}\vert \bm{\nu},\bm{\theta},\bm{\delta})^{\delta_{i}}\right\rbrace f(\bm{\nu}\vert \bm{\theta})f(\bm{\theta})d\bm{\nu}d\bm{\theta}.
\end{equation}
The ratio $m(\bm{1}_{N},\textbf{y})/m(\bm{\delta},\textbf{y})$ is similar to a likelihood ratio. The distribution $m(\bm{1}_{N}, \textbf{y})$ is the marginal distribution of the data associated with the full parametric model, and $m(\bm{\delta}, \textbf{y}_{\delta})$ is the marginal distribution of $\textbf{y}_{\delta}$ from the full parametric model. In (\ref{subhmnotation}), we have implicitly assumed $0<m(\bm{\delta},\textbf{y})<\infty$.

A key goal of our model is to address computational considerations in the Bayesian context \textit{without} making changes to the models for the processes, parameters, and the data. Our approach to do this is to introduce $\bm{\delta}$, which should be considered as a ``modeler induced error'' term. That is, the variability that arises from $\bm{\delta}$ is due to how the modeler uses the data (i.e., subsamples the data), and not because of the behavior of the processes, parameters, or the data itself. It is clear the process and parameter models have not changed between (\ref{hmnotation}) and (\ref{subhmnotation}); however, it is not immediately clear how the distribution of the data is left invariant of $\bm{\delta}$. This leads us to the following result.\\

\noindent
\textit{Proposition 1: Assume for every $\bm{\delta}$, $1\le n\le N$, and $0<m(\bm{\delta},\textbf{y})<\infty$. Then the marginal distribution of (\ref{hmnotation}) and (\ref{subhmnotation}) (across $\bm{\nu}$, $\bm{\theta}$, and $\bm{\delta}$) are equivalent and is given by $m(\bm{1}_{N},\textbf{y})$. Additionally, the model in (\ref{subhmnotation}) is proper provided that the model in (\ref{hmnotation}) is proper.}\\
\noindent
\textit{Proof:} See the Appendix.\\

\noindent
Thus, Proposition 1 shows that reweighting $\left\lbrace\prod_{i = 1}^{N} f(Y_{i}\vert \bm{\nu},\bm{\theta},\bm{\delta})^{\delta_{i}}\right\rbrace$ with the likelihood ratio $m(\bm{1}_{N},\textbf{y})/m(\bm{\delta},\textbf{y})$ is particularly important because it allows one to preserve the marginal distribution of the entire dataset $\textbf{y}$, which aids in our interpretation of $\bm{\delta}$ as modeler induced error. 

It is important to emphasize that the model in (\ref{subhmnotation}) has different properties than (\ref{hmnotation}), even though $f(\bm{\nu}\vert \bm{\theta})$, $f(\bm{\theta})$ and $m(\bm{1}_{N},\textbf{y})$ are the same in both the full and reduced parametric models.\\

\noindent
\textit{Proposition 2:}  Assume for every $\bm{\delta}$, $0<m(\bm{\delta},\textbf{y})<\infty$, $1\le n\le N$, and $0<\int m(\bm{1}_{N},\textbf{y}) d\textbf{y}_{-\delta}<\infty$, where $\textbf{y}_{-\delta} = (Y_{i}: \delta_{i} = 0)^{\prime}$. Suppose $\textbf{y}$ is distributed according to the data subset model in (\ref{subhmnotation}). Then $\textbf{y}_{\delta}$ is a partially sufficient statistic for $\bm{\nu}$ and $\bm{\theta}$.\\
\noindent
\textit{Proof:} See the Appendix.\\

\noindent
Proposition 2 states that the proposed model in (\ref{subhmnotation}) implies that $\textbf{y}_{\delta}$ is a (partially) sufficient statistic. Recall that a partially sufficient statistic $T(\textbf{y})$ for a generic real-valued transformation $T(\cdot)$ is one in which the distribution,
\begin{equation}\label{partial}
f(\textbf{y}\vert \bm{\nu},\bm{\theta},\bm{\delta}) = h(T(\textbf{y}),\bm{\nu},\bm{\theta},\bm{\delta})g(\textbf{y}, \bm{\delta}),
\end{equation}
where $h(T(\textbf{y}),\bm{\nu},\bm{\theta},\bm{\delta})$ is a non-degenerate real-valued function of $T(\textbf{y})$, $\bm{\nu}$, $\bm{\theta},$ and $\bm{\delta}$. Also, $g(T(\textbf{y}),\bm{\delta})$ is a non-degenerate real-valued function of $T(\textbf{y})$ and $\bm{\delta}$. The right-hand-side of (\ref{partial}) depends on parameters $\bm{\delta}$ but not $\bm{\nu}$ and $\bm{\theta}$. Note if $g(T(\textbf{y}),\bm{\delta})$ is constant over $\bm{\delta}$ then $T(\textbf{y})$ would be sufficient and not partially sufficient. Thus, we do not make any assumptions on $f(\bm{\nu}\vert \bm{\theta})$, $f(\bm{\theta})$, and $m(\bm{1}_{N},\textbf{y})$, but we modify the relationship between the unknowns $\{\bm{\nu},\bm{\theta}\}$ and the data $\textbf{y}$ (i.e., partial sufficiency).

\subsection{Bayesian Implementation of the Reduced Model} The posterior distribution associated with the reduced parametric model in (\ref{subhmnotation}) is given by
\begin{equation}\label{posterior}
f_{DSM}(\bm{\nu},\bm{\theta},\bm{\delta}\vert \textbf{y},n) \equiv f(\bm{\nu},\bm{\theta}\vert \textbf{y}, \bm{\delta}, n) Pr(\bm{\delta}\vert \textbf{y},n)=f(\bm{\nu},\bm{\theta}\vert \textbf{y}, \bm{\delta}, n) Pr(\bm{\delta}\vert n),
\end{equation}
\noindent
where ``DSM'' stands for ``data subset model'' and the equality in (\ref{posterior}) holds as a consequence of Proposition 3.\\

\noindent
\textit{Proposition 3:} Assume the model in (\ref{subhmnotation}) holds where for every $\bm{\delta}$, $1 \le n \le N$, and $0<m(\bm{\delta},\textbf{y})<\infty$. Then $\textbf{y}$ and $\bm{\delta}$ are independent.\\
\noindent
\textit{Proof:} See the Appendix.\\

\noindent
Proposition 3 is useful from a computational perspective. That is, we are considering the setting where $N$ is so large that it is not practical to use the entire dataset. If Proposition 3 did not hold then one would have to learn the value of $\bm{\delta}$ using the entire $N$-dimensional dataset $\textbf{y}$, which would create another computational issue. Moreover, it is crucial for our interpretation of $\bm{\delta}$ as a ``modeler induced error'' term, since dependence would suggest that the variability induced by $\bm{\delta}$ may occur due to the properties of the dataset $\textbf{y}$.

 Simulating from the posterior distribution in (\ref{posterior}) can be done efficiently using a composite sampler \citep{gelfand1990sampling}. That is,
\begin{enumerate}\setcounter{enumi}{-1}
	\item Set $g = 0$ and choose an initialization for $\bm{\delta}^{[0]}$, $\bm{\nu}^{[0]}$, and $\bm{\theta}^{[0]}$
	\item First sample $\bm{\delta}^{[g]}$ from $Pr(\bm{\delta}\vert n)$.
	\item Then, sample $\bm{\nu}^{[g]}$ and $\bm{\theta}^{[g]}$ from $f(\bm{\nu},\bm{\theta}\vert \textbf{y}, \bm{\delta}, n)$, where
	\begin{equation}
	\nonumber
	f(\bm{\nu},\bm{\theta}\vert \textbf{y}, \bm{\delta}, n)\propto\left\lbrace\prod_{\{i: \delta_{i} = 1\}}f(Y_{i}\vert \bm{\nu},\bm{\theta},\bm{\delta}) \right\rbrace f(\bm{\nu}\vert \bm{\theta})f(\bm{\theta}).
	\end{equation}
	\item Set $g = g+1$
	\item Repeat Steps 1 $\--$ 3 until $g = G$.
\end{enumerate}
Steps 1 and 2 show that at each step of the composite sampler we subset the data when updating $\bm{\nu}$ and $\bm{\theta}$. However, we include \textit{all} the data in our analysis because a new subset is generated in Step 1. Thus, we obtain the same computational benefits of using a single subset of the dataset while using the entire dataset. Similar sampling schemes exist in the machine learning literature \citep[e.g., see][among others]{kleiner2012big,Bardenet,korattikara2014austerity}, however, these approaches treat Steps 1 and 2 as a means to approximately sample from the posterior distribution of $\bm{\nu}$ and $\bm{\theta}$ using the full parametric model. Our perspective is that the reduced parametric model is simply different from the full parametric model, is an exact model, and uses the entire dataset.

\subsection{A Semi-Parametric Interpretation of the Reduced Model's Posterior Distribution}  We do not make any assumptions on $f(\bm{\nu}\vert \bm{\theta})$, $f(\bm{\theta})$, and $m(\bm{1}_{N},\textbf{y})$, but we modify the relationship between the unknowns $\{\bm{\nu},\bm{\theta}\}$ and the data $\textbf{y}$ (i.e., partial sufficiency). Moreover, the results in this section suggest that posterior inference from the reduced parametric model removes assumptions on the full parametric model. That is, suppose we drop assumptions on the full parametric model stated in (\ref{hmnotation}):
\begin{align}
\nonumber
&\mathrm{Data\hspace{5pt}Model:}\hspace{5pt} w_{\delta}(\textbf{y}_{-\delta})\underset{\{i: \delta_{i}=1\}}{\prod} f(Y_{i} \vert \bm{\nu},\bm{\theta},\bm{\delta})\\
\nonumber
&\mathrm{Process\hspace{5pt}Model:}\hspace{5pt} f(\bm{\nu}\vert \bm{\theta})\\
\nonumber
&\mathrm{Parameter\hspace{5pt}Model:} \hspace{5pt}f(\bm{\theta})\\
\label{sphmnotation}
&\mathrm{Subset \hspace{5pt}Model:} \hspace{5pt}Pr(\bm{\delta}\vert n).
\end{align}
\noindent
The density $w_{\delta}(\cdot)$ is the true unknown unparameterized probability density function for the $(N - n)$-dimensional random vector $\textbf{y}_{-\delta}$ for a given $\bm{\delta}$. We refer to the model in (\ref{sphmnotation}) as the semi-parametric full model (SFM).

Implicitly we are assuming conditional independence between the $n$-dimensional vector $\textbf{y}_{\delta}$ and the $(N-n)$-dimensional vector $\textbf{y}_{-\delta}$, given $\bm{\nu}$ and $\bm{\theta}$. This assumption is typical in Bayesian hierarchical modeling, as $\bm{\nu}$ and $\bm{\theta}$ are often interpreted as values that induce dependence among the observations \citep[e.g., see][for discussions]{cressie-wikle-book, banerjee-etal-2004}. Additionally, we assume independence between $\textbf{y}_{-\delta}$ and $\bm{\nu}$ and $\bm{\theta}$. This is reasonable considering that $w_{\delta}(\cdot)$ is specified to be the true unparameterized probability density function, which contains no parameters in its expression. 

The model in (\ref{sphmnotation}) is important for two reasons. The first, is that the semi-parametric full model is more general than the full parametric model and implies partial sufficiency of $\textbf{y}_{\delta}$. Note that the semi-parametric full model is more general than the full parametric model, since replacing $w_{\delta}(\textbf{y}_{-\delta})$ with $\underset{\{i: \delta_{i}=0\}}{\prod} f(Y_{i} \vert \bm{\nu},\bm{\theta})$ reproduces the full parametric model. \\

\noindent
\textit{Proposition 4:} Suppose the model in (\ref{sphmnotation}) holds, $1 \le n\le N$, and $0<w_{\delta}(\textbf{y}_{-\delta})<\infty$. Then $\textbf{y}_{\delta}$ is a partially sufficient statistic for $\bm{\nu}$ and $\bm{\theta}$.\\
\textit{Proof:} See the Appendix.\\

\noindent
The fact that the semi-parametric full model is more general than the full parametric model, and has this partial sufficiency property (as stated in Proposition 4), gives credence to partial sufficiency of the reduced more (as stated in Proposition 2). However, there is a more direct between the semi-parametric full model and the reduced parametric model.\\

\noindent
\textit{Proposition 5:} Suppose the model in (\ref{sphmnotation}) holds and $0<m(\bm{\delta}, \textbf{y})<\infty$ holds almost surely. Let $f_{DSM}(\bm{\nu},\bm{\theta}\vert \textbf{y},\bm{\delta},n)$ and $f_{SFM}(\bm{\nu},\bm{\theta}\vert \textbf{y},\bm{\delta},n)$ be the conditional distribution of $\bm{\nu}$ and $\bm{\theta}$ given $\textbf{y}$, $\bm{\delta}$, and $n$ under the model in (\ref{subhmnotation}) and (\ref{sphmnotation}), respectively. Then $f_{DSM}(\bm{\nu},\bm{\theta}\vert \textbf{y},\bm{\delta},n) = f_{SFM}(\bm{\nu},\bm{\theta}\vert \textbf{y},\bm{\delta},n)$.\\
\textit{Proof:} See the Appendix.\\

\noindent
The relationship between the reduced parametric model and the semi-parametric full model in Proposition 5 arises from the fact that the reduced parametric model does not require parametric specifications of $\textbf{y}_{-\delta}$ for a given $\bm{\delta}$. Furthermore, an equivalence exists between the posterior distributions of the semi-parametric full model and the reduced parametric model under an additional assumption.\\

\noindent
\textit{Proposition 6:} Suppose the model in (\ref{sphmnotation}) holds, $1 \le n \le N$, and $0<m(\bm{\delta}, \textbf{y})<\infty$. Let $f_{DSM}(\bm{\nu},\bm{\theta}\vert \textbf{y},n)$ and $f_{SFM}(\bm{\nu},\bm{\theta}\vert \textbf{y},,n)$ be the posterior distribution of $\bm{\nu}$ and $\bm{\theta}$ under the model in (\ref{subhmnotation}) and (\ref{sphmnotation}), respectively. Assume that $\textbf{y}$ and $\bm{\delta}$ are independent. Then $f_{DSM}(\bm{\nu},\bm{\theta}\vert \textbf{y}, n) = f_{SFM}(\bm{\nu},\bm{\theta}\vert \textbf{y}, n)$.\\
\textit{Proof:} See the Appendix.\\

\noindent
As discussed below Proposition 3, there are computational and conceptual reasons why this assumption is reasonable. However, we recognize that extending this approach to informative sampling is a worthwhile endeavor. To do this, estimates/expressions of 
\begin{equation*}
P(\bm{\delta}\vert \textbf{y}, n) = \frac{w_{\delta}(\textbf{y}_{-\delta})m(\bm{\delta},\textbf{y})Pr(\bm{\delta}\vert n)}{\underset{\bm{\delta}\in \Delta}{\sum}w_{\delta}(\textbf{y}_{-\delta})m(\bm{\delta},\textbf{y})Pr(\bm{\delta}\vert n)},
\end{equation*}
\noindent
are required, where $\Delta$ is the set of all possible values of $\bm{\delta}$.

Proposition 6 is crucial for illustrating that we have obtained our goal described in the introduction. Specifically, the reduced parametric model does not alter the assumptions governing $f(\bm{\nu}\vert \bm{\theta})$, $f(\bm{\theta})$, and $m(\bm{1}_{N},\textbf{y})$. Furthermore, posterior inference from the reduced parametric model does not place more restrictive assumptions on the full parametric model, but rather, \textit{removes} assumptions on the full parametric model. This is because the reduced parametric model's posterior distribution is equivalent to the semi-parametric full model's posterior distribution (as stated in Proposition 6).

Proposition 6 also provides another interpretation of $n$ in the reduced model. That is, the posterior distribution of the reduced parametric model is derived from a purely nonparametric model (and the data and the processes/priors are completely independent) as the subset size $n$ decreases to zero. Also, when $n = N$ we obtain the full parametric model, which is a purely parametric Bayesian model.
%

\begin{algorithm}[htp]\caption{Algorithm 1: Implementing the model in Section 2.3.}\label{euclid}
	\begin{algorithmic}[1]
		\item  {\small Specify a sequence of values for $n$, and denote it with $n_{1},\ldots, n_{B}$. For each choice of $n_{i}$ compute Steps 2 through 14 in parallel. }
		\vfill
		\item {\small Set $g = 1$, $h = 0$, and initialize $\bm{\delta}^{[0]}$, $\bm{\beta}^{[0]}$, $\bm{\eta}^{[0]}$, $\bm{\xi}^{[0]} = \left(\xi_{1}^{[0]},\ldots, \xi_{N}^{[0]}\right)^{\prime}$, $\sigma^{2[0]}$, $\sigma_{\xi}^{2[0]}$, and $\sigma_{\beta}^{2[0]}$. Define a set of predictions $A\subset \{1,\ldots, N\}$, where $A$ contains $m\ll N$ values.}
		\vfill
		\item {\small Sample $\bm{\delta}^{[g]} = (\delta_{1}^{[g]},\ldots, \delta_{N}^{[g]})^{\prime}$ from $Pr(\bm{\delta}\vert n)$. Let the $n\times g$ matrix $\textbf{X}_{\delta} = \left(\textbf{x}_{i}: \delta_{i} = 1\right)^{\prime}$, and the $n\times r$ matrix $\bm{\Psi}_{\delta} = \left(\bm{\psi}_{i}: \delta_{i} = 1\right)^{\prime}$.}
		\item {\small Sample $\bm{\eta}_{\delta}^{[g]} = \left(\eta_{i}^{[g]}: \delta_{i} = 1\right)^{\prime}$ from it's full-conditional distribution}
		\begin{equation*}
		Normal\left\lbrace \left(\bm{\Psi}_{\delta}^{\top}\bm{\Psi}_{\delta}+ \frac{\sigma^{2[g-1]}}{\sigma_{\eta}^{2[g-1]}}\textbf{I}_{r}\right)^{-1}\bm{\Psi}_{\delta}^{\top}(\textbf{y}_{\delta} - \textbf{X}_{\delta}\bm{\beta}^{[g-1]} - \bm{\xi}_{\delta}^{[g-1]}), \left(\frac{1}{\sigma^{2[g-1]}}\bm{\Psi}_{\delta}^{\top}\bm{\Psi}_{\delta}+ \frac{1}{\sigma_{\eta}^{2[g-1]}}\textbf{I}_{r}\right)^{-1}\right\rbrace,
		\end{equation*}
		{\small where $\bm{\xi}_{\delta}^{[g-1]} = \left(\xi_{i}^{[g-1]}: \delta_{i} = 1\right)^{\prime}$.}
		\vfill
		\item {\small Sample $\bm{\xi}_{\delta}^{[g]}$ from }
		$Normal\left\lbrace \left(\frac{\sigma_{\xi}^{2[g-1]}}{\sigma^{2[g-1]}+\sigma_{\xi}^{2[g-1]}}\right)(\textbf{y}_{\delta} - \textbf{X}_{\delta}\bm{\beta}^{[g-1]} - \bm{\Psi}_{\delta}\bm{\eta}^{[g]}),\left(\frac{\sigma^{2[g-1]}\sigma_{\xi}^{2[g-1]}}{\sigma^{2[g-1]}+\sigma_{\xi}^{2[g-1]}}\right)\textbf{I}_{n}\right\rbrace.$
		\vfill
		\item {\small Sample $\bm{\beta}^{[g]}$ from it's full-conditional distribution}
		\begin{equation*}
		Normal\left\lbrace \left(\textbf{X}_{\delta}^{\top}\textbf{X}_{\delta}+ \frac{\sigma^{2[g-1]}}{\sigma_{\beta}^{2[g-1]}}\textbf{I}_{g}\right)^{-1}\textbf{X}_{\delta}^{\top}(\textbf{y}_{\delta} - \bm{\Psi}_{\delta}\bm{\eta}^{[g]} - \bm{\xi}_{\delta}^{[g]}), \left(\frac{1}{\sigma^{2[g-1]}}\textbf{X}_{\delta}^{\top}\textbf{X}_{\delta}+ \frac{1}{\sigma_{\beta}^{2[g-1]}}\textbf{I}_{g}\right)^{-1}\right\rbrace.
		\end{equation*}
		\vfill
		\item {\small Sample $\sigma^{2[g]}$ from it's full-conditional distribution }
		\begin{equation*}
		IG\left(1+\frac{n}{2},1+\frac{(\textbf{y}_{\delta} -\textbf{X}_{\delta}\bm{\beta}^{[g]}- \bm{\Psi}_{\delta}\bm{\eta}^{[g]} - \bm{\xi}_{\delta}^{[g]})^{\top}(\textbf{y}_{\delta} -\textbf{X}_{\delta}\bm{\beta}^{[g]}- \bm{\Psi}_{\delta}\bm{\eta}^{[g]} - \bm{\xi}_{\delta}^{[g]})}{2}\right).
		\end{equation*}
		\vfill
		\item {\small Sample $\sigma_{\eta}^{2[g]}$ from it's full-conditional distribution} $IG\left(1+\frac{n}{2},1+\frac{\bm{\eta}_{\delta}^{[g]\top}\bm{\eta}_{\delta}^{[g]}}{2}\right)$.
		\vfill
		\item {\small Sample $\sigma_{\xi}^{2[g]}$ from it's full-conditional distribution $IG\left(1+\frac{n}{2},1+\frac{\bm{\xi}_{\delta}^{[g]\top}\bm{\xi}_{\delta}^{[g]}}{2}\right)$.} {\small Sample $\sigma_{\beta}^{2[g]}$ from it's full-conditional distribution $IG\left(1+\frac{p}{2},1+\frac{\bm{\beta}^{[g]\top}\bm{\beta}^{[g]}}{2}\right)$.} 
		\item {\small Sample $\bm{\eta}_{-A}^{[g]} = \left(\eta_{i}^{[g]}: \delta_{i} = 0, i \in A\right)^{\prime}$ from it's full-conditional distribution $Normal(\bm{0}_{m}, \sigma_{\eta}^{2[g-1]})$.}
		\vfill
		\item {\small Sample $\bm{\xi}_{-A}^{[g]} = \left(\xi_{i}^{[g]}: \delta_{i} = 0, i \in A\right)^{\prime}$ from it's full-conditional distribution $Normal(\bm{0}_{m}, \sigma_{\xi}^{2[g-1]})$.}
		\vfill
		\item {\small For each $i \in A$ compute $\mu_{i}^{[g]} = \textbf{x}_{i}^{\prime}\bm{\beta}^{[g]}+\bm{\psi}_{i}^{\prime}\textbf{D}_{A}\bm{\eta}^{[g]}+\xi_{i}^{[g]}$, where $\textbf{D}_{A} = diag\left\lbrace I(i \in A): i = 1,\ldots, N\right\rbrace$.}
		\vfill
		\item {\small Let $g = g+1$.}
		\vfill
		\item {\small Repeat Steps 3 through 13 until $g = G$.}
		\vfill
		\item {\small For each $n_{i}$ record the CPU time needed for Steps 2$\--$14.}
	\end{algorithmic}
\end{algorithm}
\subsection{Example Model: A Bayesian Hierarchical with a data subset model for Normal Data}
Assume that the data $Y_{1},\ldots, Y_{N}$ are distributed normally. That is, consider the following specification of $f(Y_{i}\vert \bm{\nu},\bm{\theta},\bm{\delta})$:
\begin{align*}
Y_{i}\vert \bm{\nu},\bm{\theta},\bm{\delta}\sim Normal(\textbf{x}_{i}^{\prime}\bm{\beta}+\bm{\psi}_{i}^{\prime}\textbf{D}_{\delta}\bm{\eta} + \xi_{i}),
\end{align*}
where $Normal(\mu,v)$ is a shorthand for the normal density with mean $\mu$ and covariance $v>0$. Let $\textbf{x}_{i}$ be a known $g$-dimensional covariates,	$\bm{\psi}_{i}$ be a known $N$-dimensional vector of basis functions, $\textbf{D}_{\delta}$ be a $N\times N$ diagonal matrix with $\bm{\delta}$ on the main diagonal. For every $\bm{\delta}$ the $n$-dimensional vector $(Y_{i}: \delta_{i} = 1)^{\prime}$ is modeled with a full rank basis function expansion, which is a common model choice in functional analysis \citep{wahba}, nonparametric Bayesian analysis \citep{barry1986nonparametric}, and spatial statistics \citep{wikleHandbook}. Let $\bm{\eta}$ be $N$-dimensional, $\bm{\xi}=(\xi_{1},\ldots, \xi_{N})^{\prime}$ be $N$-dimensional, $\bm{\eta}\vert\sigma^{2}\sim Gau(\bm{0},\sigma^{2}\textbf{I}_{N})$, $\textbf{I}_{N}$ is a $N\times N$ identity matrix, $\bm{\xi}\vert\sigma_{\xi}^{2}\sim Gau(\bm{0}_{N},\sigma_{\xi}^{2}\textbf{I}_{N})$, $\bm{\beta}\vert\sigma_{\beta}^{2}\sim Normal(\bm{0}_{p},\sigma_{\beta}^{2}\textbf{I}_{p})$, $\sigma^{2}\sim IG(a,b)$, $\sigma_{\xi}^{2}\sim IG(a,b)$, $\bm{\nu}=(\bm{\eta}^{\prime},\bm{\xi}^{\prime})^{\prime}$ is $r =2N$-dimensional, $\bm{\theta}=(\bm{\beta}^{\prime},\tau^{2},\sigma^{2},\sigma_{\xi}^{2},\sigma_{\beta}^{2})^{\prime}$, and $IG(a,b)$ is a shorthand for an inverse-gamma distribution with shape $a>0$ and rate $b>0$. Finally, we specify $Pr(\bm{\delta}\vert n)$ based on random sampling without replacement \citep{lohr} so that $Pr(\delta_{i} = 1\vert n) = n/N$.  
 
 Algorithm 1 describes the implementation of the normal model described above. Essentially, we perform Items 1 and 2 from Section 2.2, where Step 2 is implemented using a Gibbs sampler (with a reversible jump), and we run the algorithm in parallel over discrete specifications of the hyperparameter $n = 1,\ldots, n_{B}$. Thus, the computation time of this algorithm is equal to the time it takes to implement Steps 2 through 14 in Algorithm 1 for $n_{B}$. When the computation time at $n = n_{B}$ is greater than 5 minutes (or some other prespecified time), we choose the value of $n$ such that the computation gets closest to 5 minutes.

\subsection{Example Model: A Bayesian Hierarchical with a  data subset model for Bernoulli Data}
Consider the following specification of $f(Y_{i}\vert \bm{\nu},\bm{\theta},\bm{\delta})$:
\begin{align*}
Y_{i}\vert \bm{\nu},\bm{\theta},\bm{\delta}\sim Bernoulli(\textbf{x}_{i}^{\prime}\bm{\beta}+\bm{\psi}_{i}^{\prime}\textbf{D}_{\delta}\bm{\eta} + \xi_{i},\tau^2),
\end{align*}
where $Bernoulli(\mu)$ is a shorthand for the Bernoulli distribution with probability of success $\mathrm{exp}(\mu)/\left\lbrace 1+\mathrm{exp}(\mu)\right\rbrace$. Here $\textbf{D}_{\delta} = \textbf{I}_{r}$ and $\bm{\psi}_{i}$ is $r$-dimensional. We model $\bm{\eta}$ with the multivariate logit-beta (MLB) distribution introduced in \citet[][]{bradleyLCM}. That is, $\bm{\eta}$ has the following probability density function (pdf) \citep{bradleyLCM, bradleyJTSA,gao2019bayesian}:

\vspace{-30pt}
\begin{align}\label{cm}
\nonumber
& f(\bm{\eta}|\bm{\mu}, \textbf{V}, \bm{\alpha}, \bm{\kappa})\\
&= \mathrm{det}(\textbf{V}^{-1})\left\lbrace \prod_{i = 1}^{n}\frac{\Gamma(\kappa_{i})}{\Gamma(\alpha_{i})\Gamma(\kappa_{i}-\alpha_{i})}\right\rbrace \mathrm{exp}\left[\bm{\alpha}^{\prime} \textbf{V}^{-1}(\bm{\eta} - \bm{\mu}) - \bm{\kappa}^{\prime}log\left\lbrace \bm{1}+ exp\left(\textbf{V}^{-1}(\bm{\eta}-\bm{\mu})\right)\right\rbrace\right],
\end{align}

\noindent where $\textbf{V}$ is an unknown $r\times r$ lower triangular matrix, ``$\mathrm{det}$'' denotes the determinant function, $\bm{\alpha} \equiv({\alpha}_{1},\cdots,{\alpha}_{n})^{\prime}$, $\bm{\kappa}\equiv ({\kappa}_{1},\cdots,{\kappa}_{n})^{\prime}$, and $\kappa_{i}>\alpha_{i}>0$ for each $i = 1,\ldots, r$. Let $\bm{\eta}\sim MLB(\bm{\mu}, \textbf{V}, \bm{\alpha}, \bm{\kappa})$ represent that $\bm{\eta}$ is distributed as a multivariate logit-beta distribution with location parameter $\bm{\mu}$, precision parameter $\textbf{V}$, shape parameter $\bm{\alpha}$ and shape parameter $\bm{\kappa}$.

Let $\bm{\xi}=(\xi_{1},\ldots, \xi_{N})^{\prime}\sim MLB(\bm{0}_{N},\textbf{I}_{N},\alpha_{\xi},\kappa_{\xi})$ be $N$-dimensional, the $r$-dimensional random vector $\bm{\eta}\vert\sigma^{2}\sim MLB(\bm{0},\textbf{V})$, $\textbf{V}$ is a Cholesky matrix whose elements are modeled independently as MLB with location zero, precision one, and fixed shape parameters, $\bm{\xi}\vert\sigma_{\xi}^{2}\sim MLB(\bm{0},\sigma_{\xi}^{2}\textbf{I})$, $\bm{\beta}\vert\sigma_{\beta}^{2}\sim MLB(\bm{0},\sigma_{\beta}^{2}\textbf{I})$, $\bm{\nu}=(\bm{\eta}^{\prime},\bm{\xi}^{\prime})^{\prime}$ is $(r+N)$-dimensional, and $\bm{\theta}=(\bm{\beta}^{\prime},\tau^{2},\sigma^{2},\sigma_{\xi}^{2},\sigma_{\beta}^{2})^{\prime}$.  Similar to the model in Section 2.3, we perform Items 1 and 2 from Section 2.2, where Step 2 is implemented using a collapsed Gibbs sampler from \citet{bradleyLCM}. We run the algorithm in parallel over discrete specifications of the hyperparameter $n = 1,\ldots, n_{B}$. When the computation time at $n = n_{B}$ is greater than 5 minutes (or some other prespecified time), we choose the value of $n$ such that the computation gets closest to 5 minutes.

\section{Illustrations}     

\subsection{Simulated Example} Generate a dataset of size $N = 100,000,000$ from an order one autoregressive model with measurement error. Specifically, let
\begin{equation*}
Y_{i} = \mu_{i} + \epsilon_{i},
\end{equation*} 
\noindent
where 
\begin{equation*}
\nu_{i} = 0.9\mu_{i-1} + \zeta_{i}.
\end{equation*}
Generate $\epsilon_{i}$ and $ \zeta_{i}$ independently from $Normal(0,0.1)$. For a given $i$, compute the following prediction:
\begin{equation*}
\widehat{\mu}_{i}=\frac{1}{G-g_{0}}\sum_{g = g_{0}+1}^{G}\mu_{i}^{[g]};\hspace{5pt} i \in A \subset \{1,\ldots, N\},
\end{equation*}
\noindent
\noindent
where $A=\{w(j): j = 1,\ldots, 100,000\} \subset \{1,\ldots, N\}$ are $100,000$ equal spaced time points over $1,\ldots, N$. Let $\widehat{\bm{\mu}}_{n} = \left(\widehat{\mu}_{w(j)}: j = 1,\ldots, 100,000\right)^{\prime}$ be the predicted values of $\{\mu_{i}\}$ using the average (across $g$) of the replicates in Step 12 of Algorithm 1. The $(i,j)$-th element of $\bm{\Psi}_{\delta}$ (see the definition in Step 3 of Algorithm 1) is chosen to be a Gaussian radial basis function:
\begin{equation*}
\mathrm{exp}\left(-\rho |i - j|\right),
\end{equation*}
\noindent
where $\rho = 0.3$ is fixed and $|\cdot|$ is the absolute value. To evaluate our method we compute the root mean squared prediction error (RMSPE):
\begin{equation*}
RMSPE = \left\lbrace \frac{1}{M}\sum_{j = 1}^{M}(\mu_{w(j)}-\widehat{\mu}_{w(j)})^{2}\right\rbrace^{1/2},
\end{equation*}
\noindent
and the central processing unit (CPU) in seconds.

\begin{figure}
	\begin{center}
	\begin{tabular}{cc}
		\includegraphics[height=3in, width=3in]{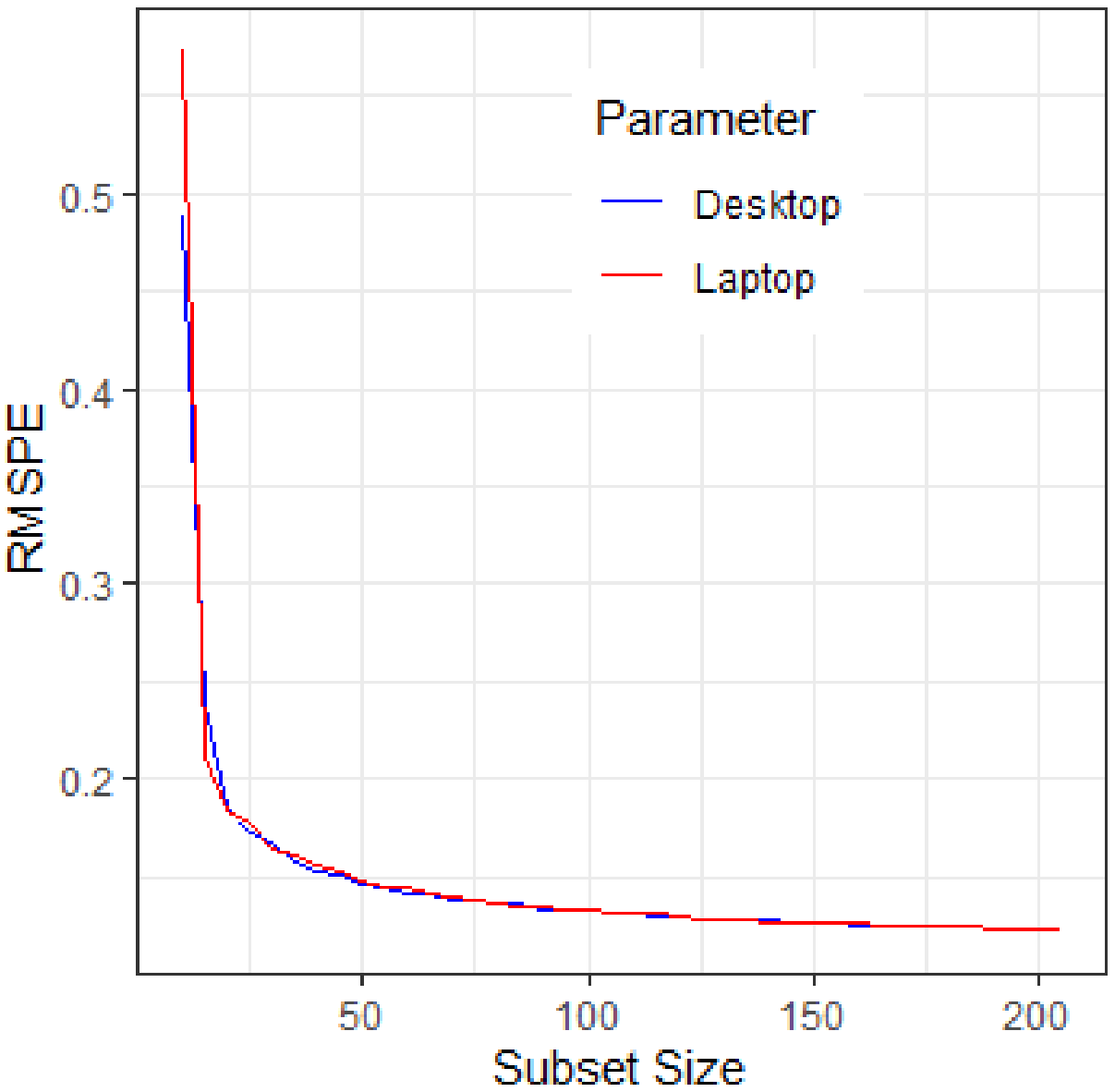}&\includegraphics[height=3in, width=3in]{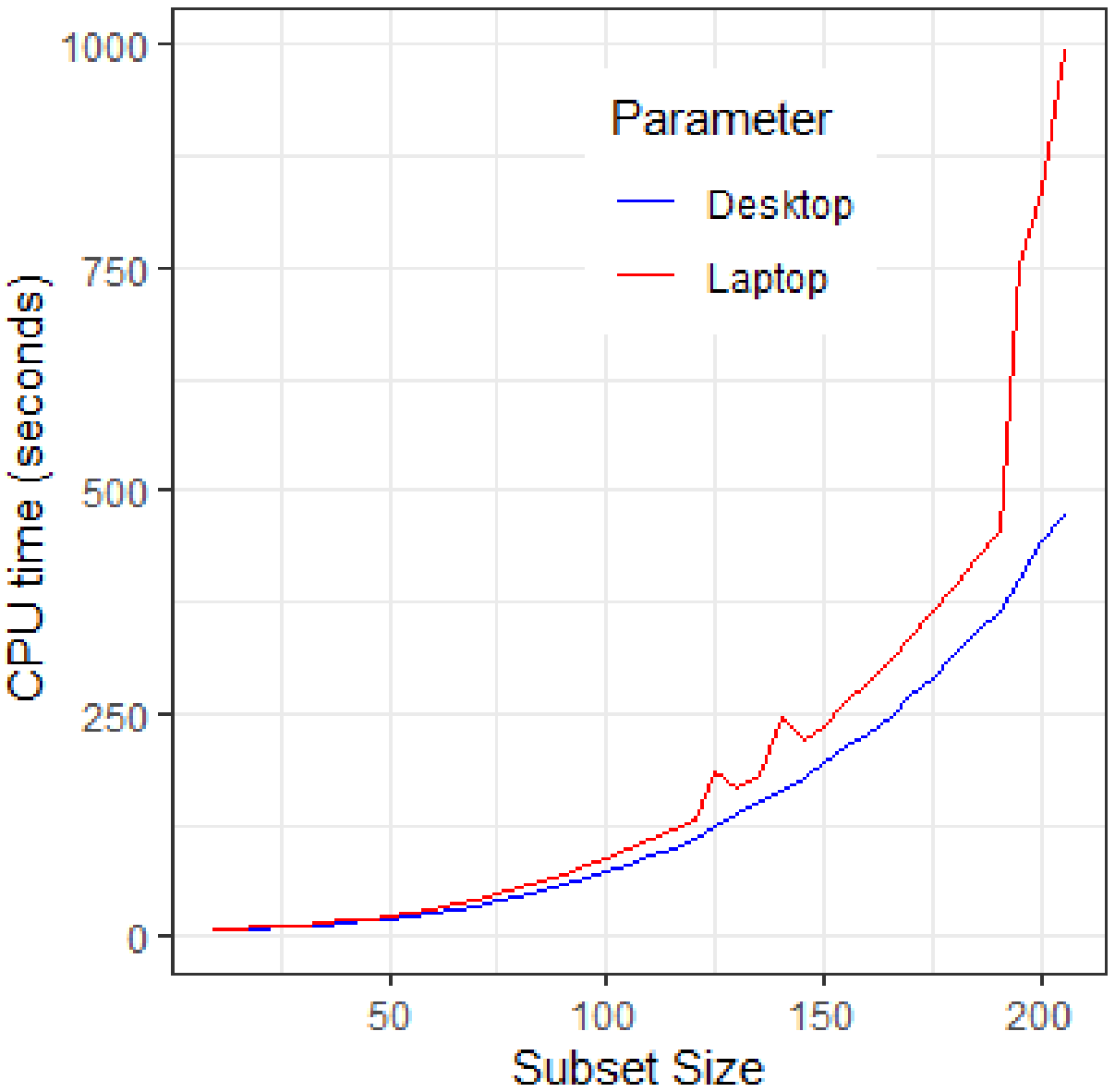}
	\end{tabular}
\end{center}
	\caption{In the left panel we plot the RMSPE by computer (desktop or laptop) and sample size. In the right panel we plot the CPU time of Algorithm 1 by computer (desktop or laptop) and sample size.}\label{fig:1}
\end{figure}

We implement the Gibbs sampler in Algorithm 1 with $G=10,000$ iterations and a burn-in of $g_{0}=1000$. Trace plots are informally checked for this simulation study with no lack of convergence detected. Simple random sampling without replacement is used to define $Pr(\bm{\delta}\vert n)$. We consider fitting our model on two different computers. The first computer is the author's desktop computer, which is running on Windows 10 with the following specifications: Intel(R) Xeon(R) CPU E5-2640 (v3) with 2.60Ghz. The second computer is the author's laptop computer, which is running on Windows 10 with the following specifications: Intel(R) CORE(TM) i5-8250U CPU  with 1.60Ghz.
\begin{figure}
	\begin{center}
		\begin{tabular}{c}
			\includegraphics[height=3in, width=4.5in]{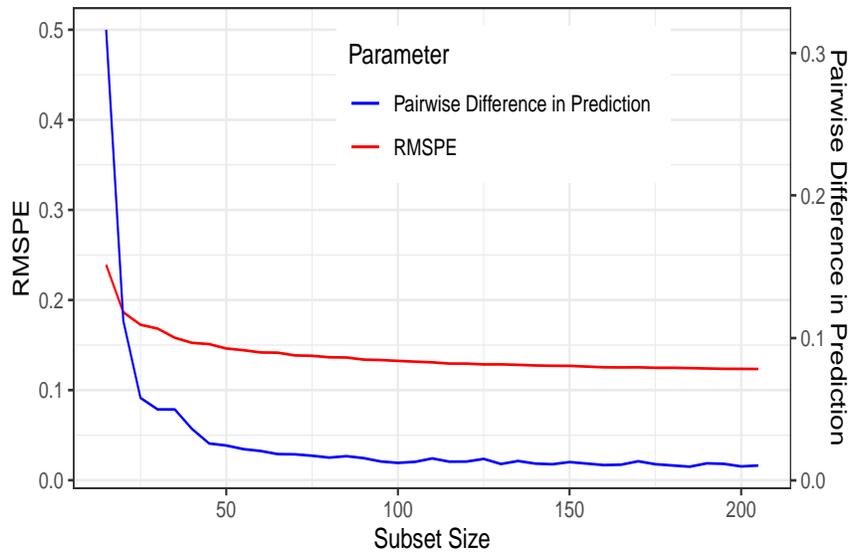}
		\end{tabular}
	\end{center}
	\caption{We plot the RMSPE and pairwise difference (i.e., $ (\widehat{\bm{\mu}}_{n}-\widehat{\bm{\mu}}_{n+1})^{\prime}(\widehat{\bm{\mu}}_{n}-\widehat{\bm{\mu}}_{n+1})$) by sample size. The left $y$-axis provides the value of RMSPE, and the right $y$-axis provides the values of the pairwise differences in the predictions.}\label{fig:2}
\end{figure}

In Figure \ref{fig:1}, we plot the RMSPE and the CPU time of Algorithm 1 by computer (desktop or laptop) and sample size. Note that spikes in CPU time are known to occur when executing lengthly loops \citep{reiss2012heterogeneity}, such as a Gibbs sampler.
In the left panel, we obtain the same prediction errors using both computers at all sample sizes, and observe an ``elbow'' pattern in the RMSPE as the sample size increases. This suggests that larger sample sizes yield smaller predictions, but the relative decrease in RMSPE is small as the sample size increases. In the right panel, we see a difference in CPU time by computer, where the CPU associated with the desktop increases at a slower rate than the CPU associated with the laptop. For the laptop a sample size $n = 162$ produces predictions in roughly 5 minutes, while for the desktop $n = 175$ produces predictions in roughly 5 minutes. Thus, the best predictor computed in 5 minutes and using the laptop has RMSPE of roughly 0.52, and the best predictor computed in 5 minutes using the desktop has RMSPE of roughly 0.50. 

Suppose we don't have a time limit, but are still not interested waiting for the model with $n = N$ to be be implemented. One solution is to choose $n$ to be a value after the ``elbow'' seen in the left panel of Figure \ref{fig:1}. However, in practice, we don't know when the RMSPE decreases at a slower rate (i.e., where the ``elbow'' occurs). Thus, we suggest computing the pairwise difference between $\widehat{\bm{\mu}}_{n}$ and $\widehat{\bm{\mu}}_{n+1}$, or $ (\widehat{\bm{\mu}}_{n}-\widehat{\bm{\mu}}_{n+1})^{\prime}(\widehat{\bm{\mu}}_{n}-\widehat{\bm{\mu}}_{n+1})$. If this value is close to zero we see little benefit in increasing the sample size. In Figure \ref{fig:2} we plot the RMSPE, which can not be computed in practice, and the pairwise difference. Notice that the ``elbow'' occurs at a similar spot for both metrics.

\begin{figure}
	\begin{tabular}{ccc}
\includegraphics[height=2.5in, width=2in]{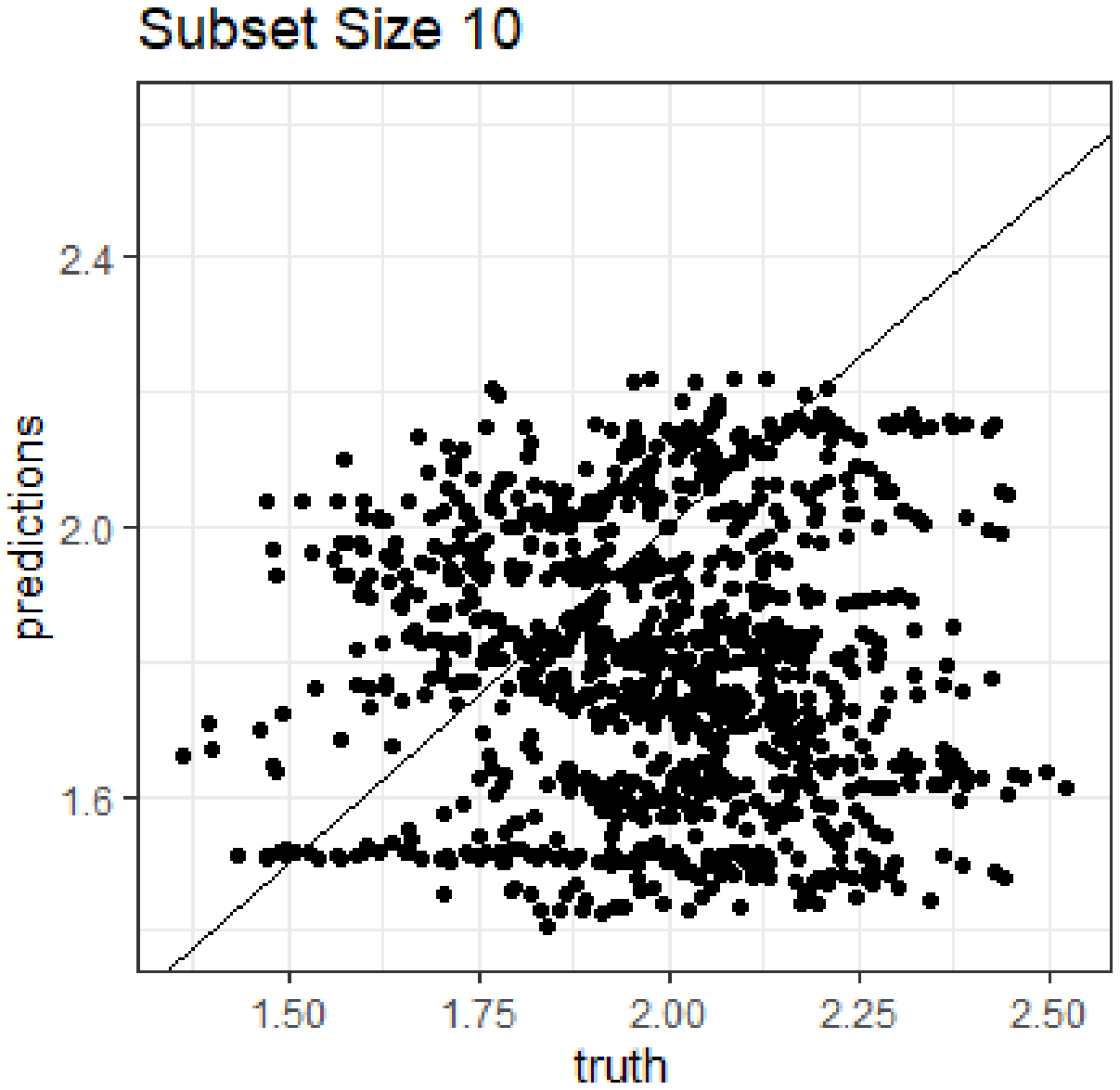} &\includegraphics[height=2.5in, width=2in]{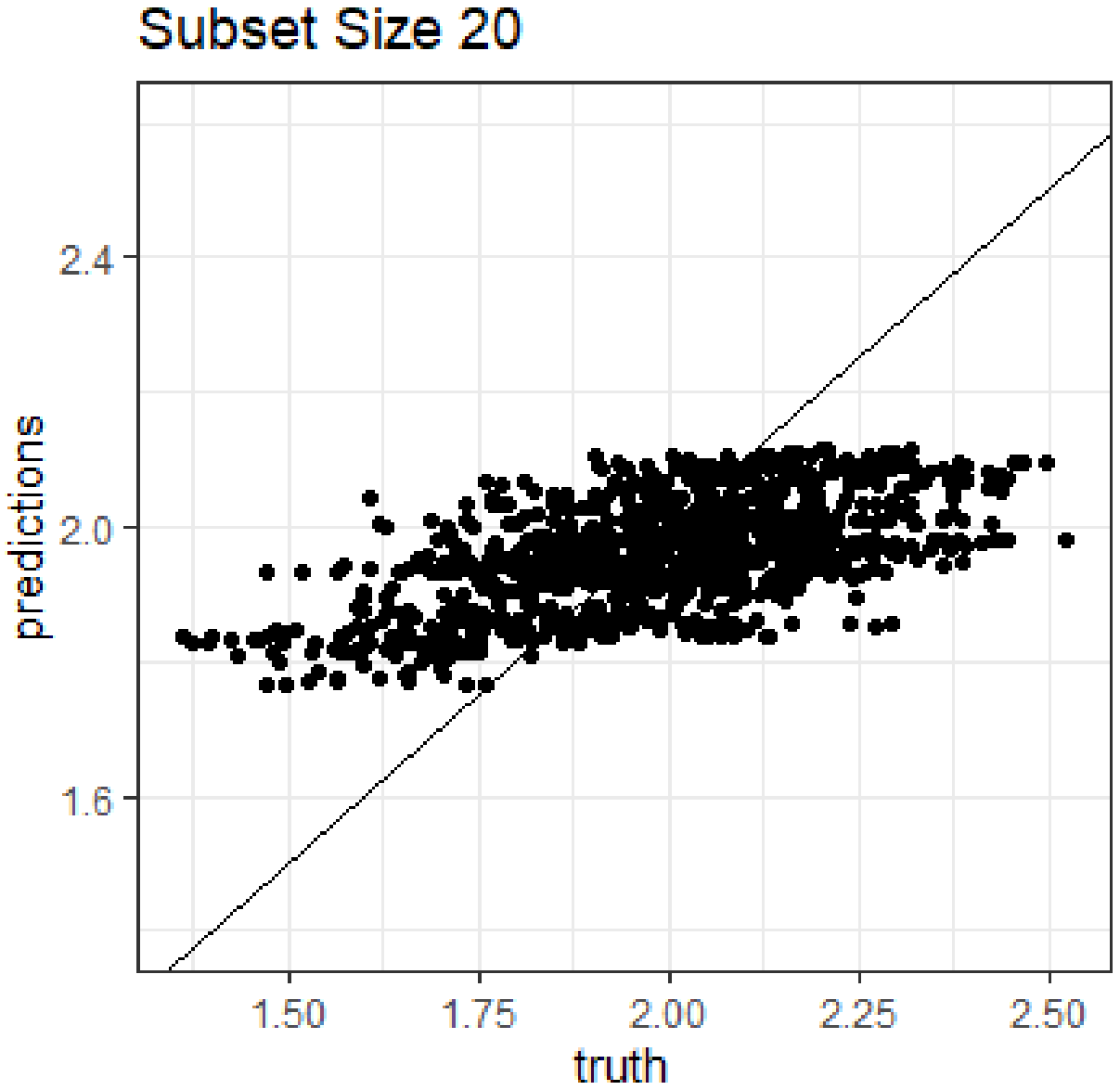}&
\includegraphics[height=2.5in, width=2in]{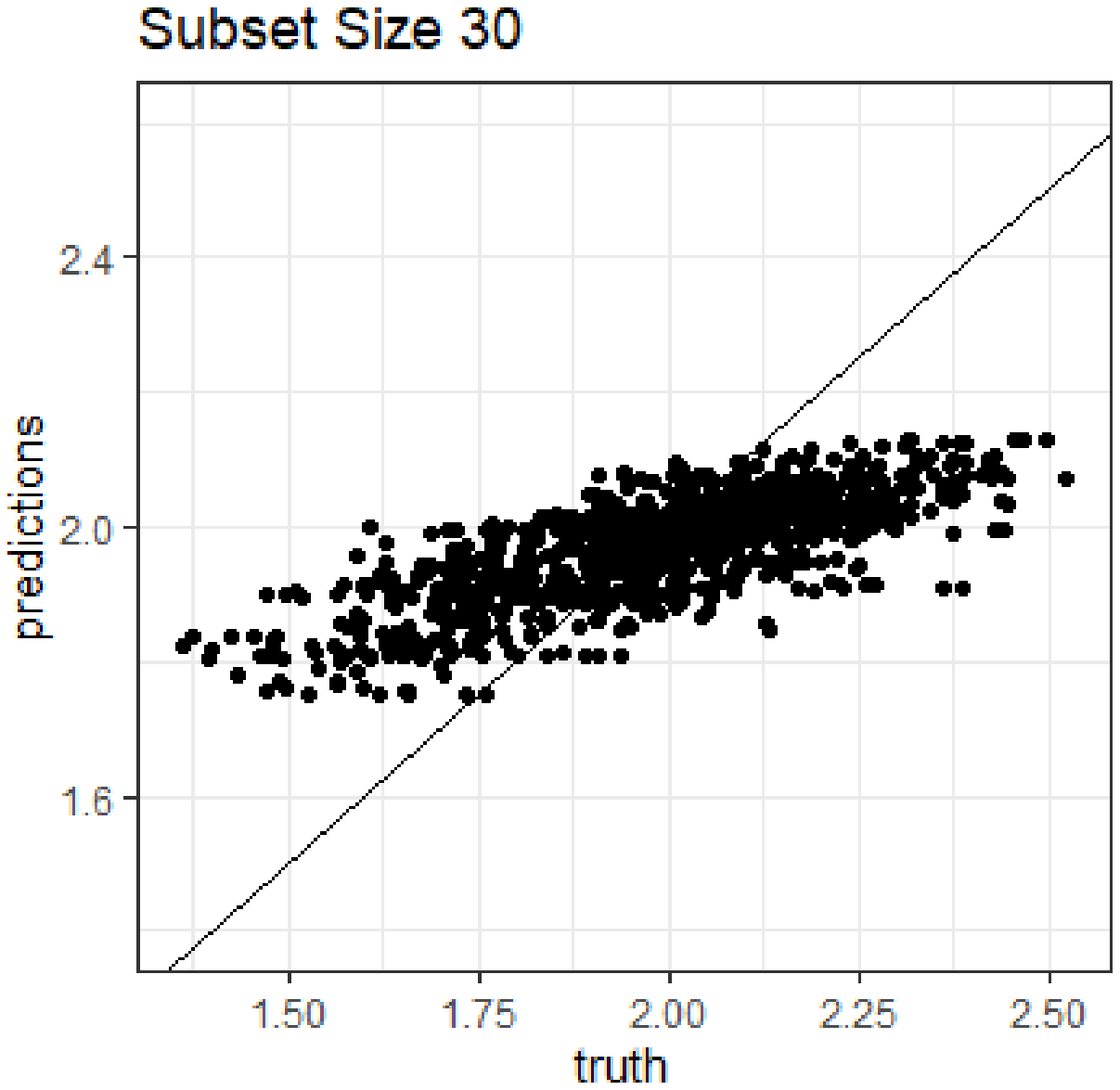}\\
\includegraphics[height=2.5in, width=2in]{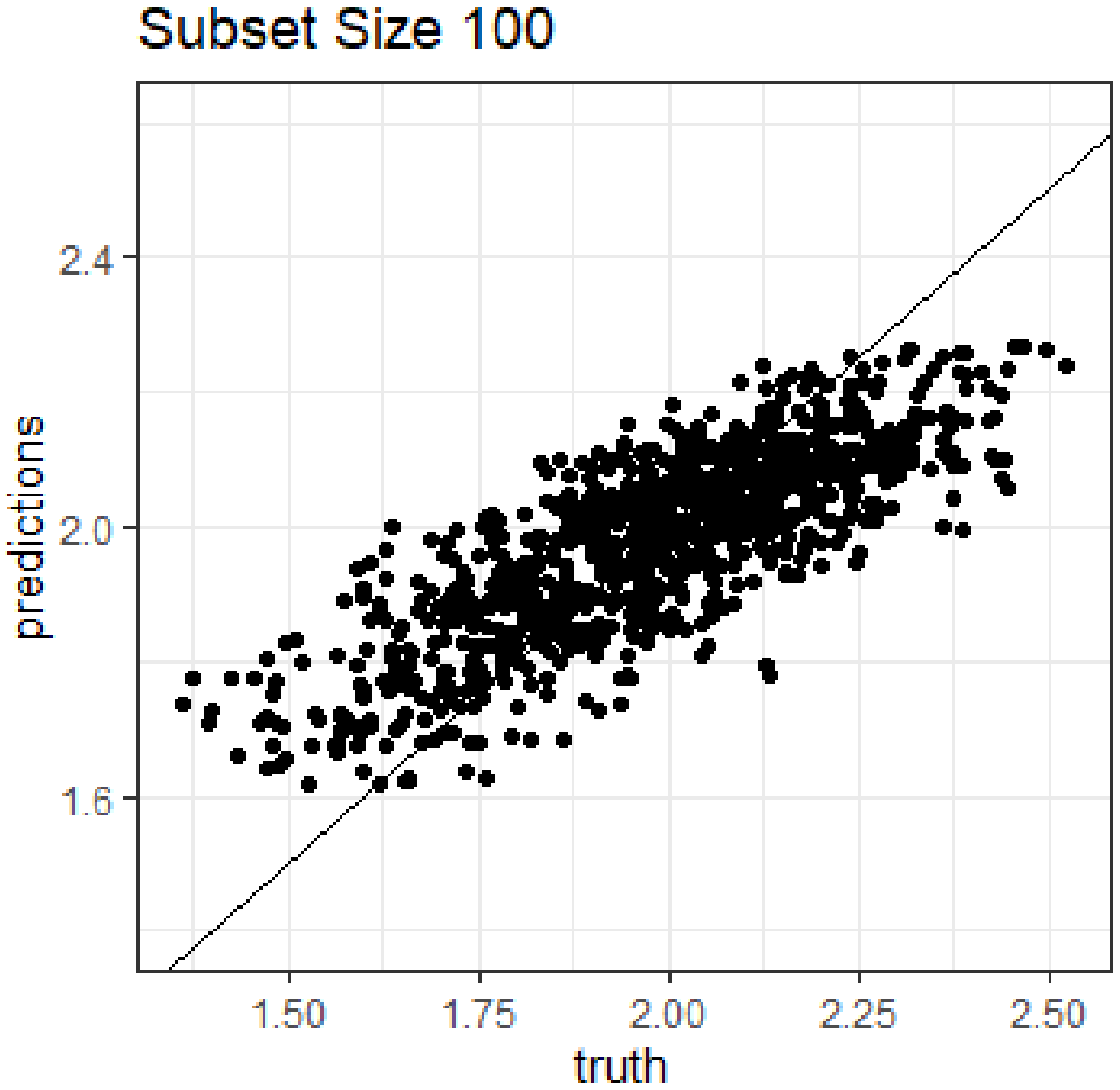}&
\includegraphics[height=2.5in, width=2in]{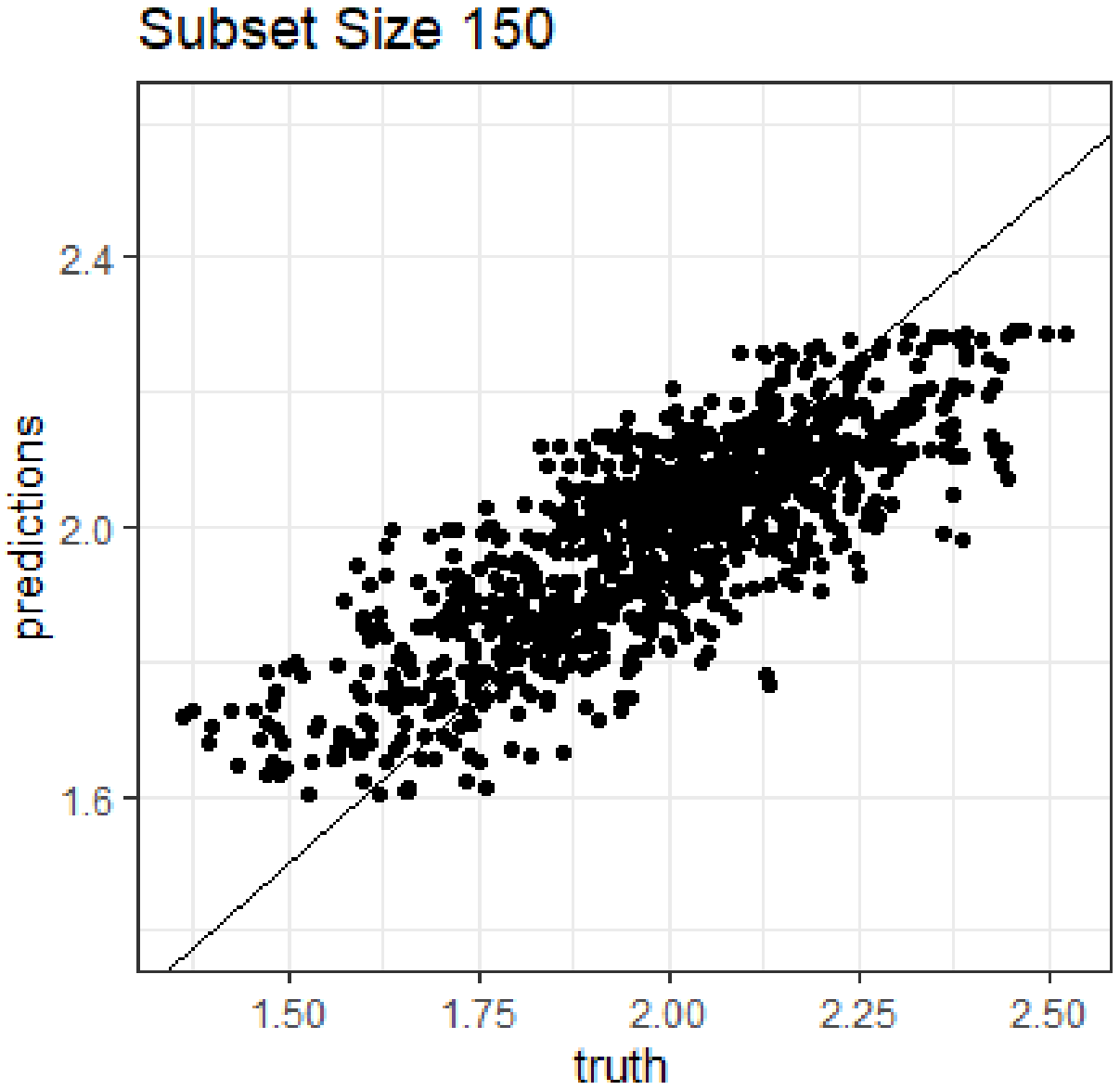}&
\includegraphics[height=2.5in, width=2in]{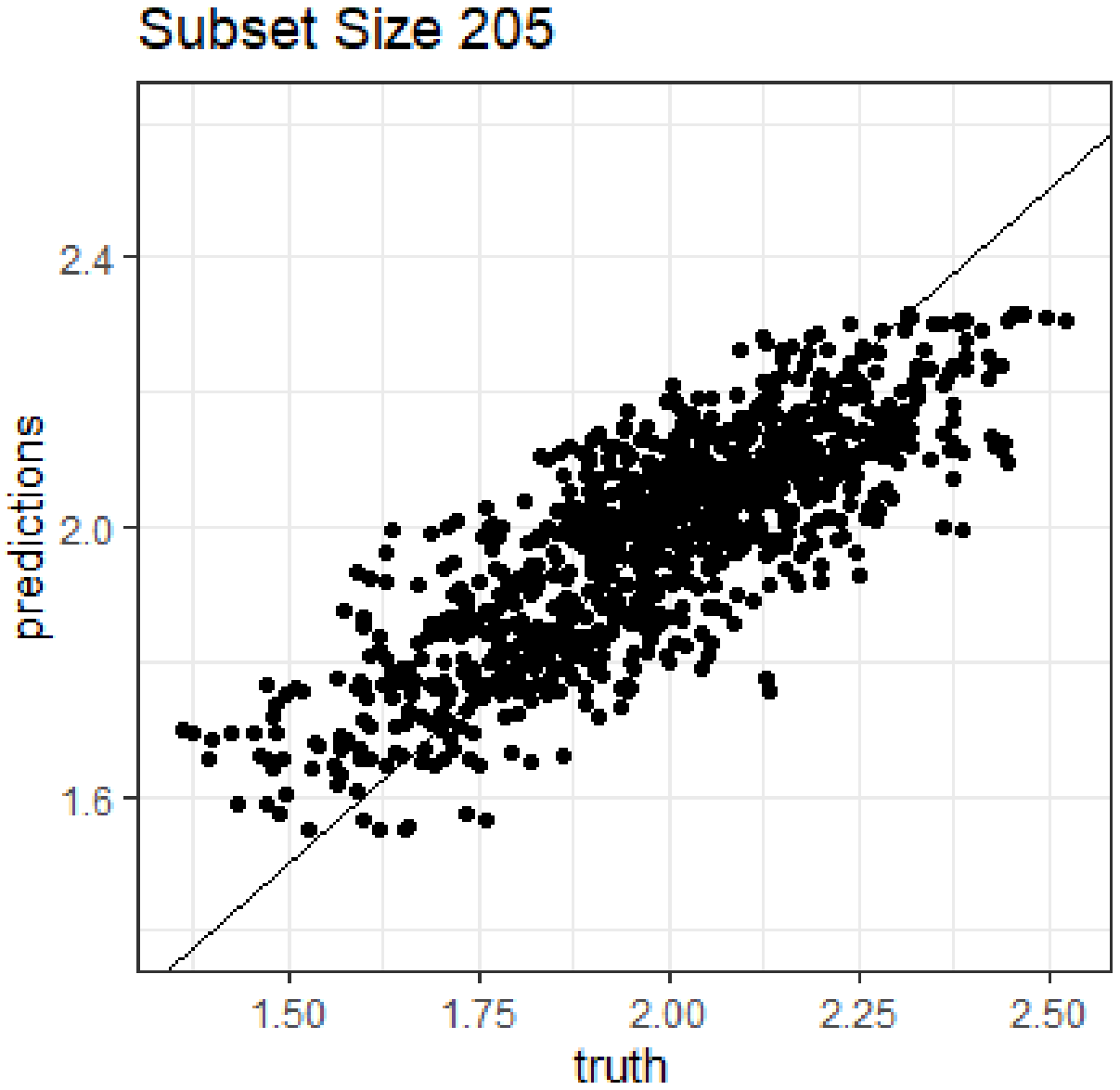}
	\end{tabular}
\caption{Plot of predicted value of $\{\mu_{i}\}$ versus the true values for $\{\mu_{i}\}$. Predicted values are computed by averaging $\mu_{i}^{[g]}$ over $g$ in Step 12 of Algorithm 1. Each panel represents these results using a different choice of $n$ as indicated in the title heading of each panel. A reference line for 45$^{\circ}$ is added}\label{fig:3}
\end{figure}

It follows intuition that, as we consider larger values of $n$, we obtain smaller prediction errors (as shown in Figure \ref{fig:1}). However, we would like to qualify how the predictions improve as $n$ increases. In Figure \ref{fig:3}, we plot the predicted value of $\{\mu_{i}\}$ versus the true values for $\{\mu_{i}\}$ for a selected subset of $\{\mu_{i}\}$. Each panel is based on a different choice of $n$ as indicated in the title heading of each panel. For small values of $n$ (i.e., $n=10,20,$ and $30$) our predictions underestimate large values and over estimate small values. In other words, the predictions look similar to the overall average. However, as $n$ increases the predictions get close to following the 45$^{\circ}$ line, which indicates reasonable predictions.

\subsection{Application: Benchmark CO$_{2}$ Dataset} We analyze a benchmark dataset introduced in \citet{bradley2014_comp}. This dataset was collected using the Atmospheric Infrared Sounder (AIRS), which is a remote-sensing instrument on board the Aqua satellite. The Aqua satellite is part of the National Aeronautics and Space Administration (NASA)'s Earth Observing System (EOS). The data represent measurements of mid-tropospheric $\mathrm{CO}_{2}$ in parts per million. We consider data from \citet{bradley2014_comp}, which were recorded from February 1 through February 9, 2010 over latitudes $\--$60$^{\circ}$ to 90$^{\circ}$. In total there are $N = 59,488$ observations, which we use for training and $14,873$ observations used for validation. \citet{bradley2014_comp} analyze $\mathrm{CO}_{2}$ in this region, and provide a comparison between several predictions including: fixed rank kriging \citep[FRK;][]{johan}, lattice kriging \citep[LTK;][]{nychkaLK}, negative exponential distance weighting (EDW), and a stochastic partial differential equation approach \citep[SPD;][]{lindgren-2011} which uses INLA. 

We implement Algorithm 1 with $n=10, 60, 110, 210,$ and $260$, $G=10,000$ iterations, a burn-in of $g_{0}=1000$, and the Gaussian radial basis function based on the spherical distance metric are used. Trace plots are informally checked, with no lack of convergence detected. Simple random sampling without replacement is used to define $Pr(\bm{\delta}\vert n)$. In Table \ref{tab:1}, we provide the root mean squared testing error,
\begin{equation*}
RSTE=\left(\frac{\sum_{i=1}^{14,873}(Y_{i}-\widehat{\mu}_{i})^{2}}{14,873}\right)^{1/2},
\end{equation*}
\noindent
where $\widehat{Y}_{i}$ represents the predicted value of $Y_{i}$ and $Y_{1},\ldots,Y_{14873}$ are the validation data. We also provide the CPU time for each method. Here we see that Algorithm 1 with $n=110$ provides a good balence between RSTE and CPU time, and we outperform all the aforementioned competing methods.

\begin{table}
\begin{center}
	\begin{tabular}{l | c | c |}
		Method & RSTE (holdout) & CPU Time (seconds)\\
		\hline \hline
		Fixed Rank Kriging & 3.9841 & 30.6\\ 
		Lindgren et al. (2011) & 3.9882 & 283.2\\
		Lattice Kriging & 4.0026 & 5,107.8\\
		Inverse Distance Weighting & 5.5203 & 16,784.4\\
		DSM, Subset 10 & 5.6522 & {\color{red}5.7}\\
		DSM, Subset 60 & 4.1088 & 39.4\\
		DSM, Subset 110 & {\color{red}3.9786} & 45.1\\
		DSM, Subset 210 & 4.0513 & 95.5\\
		DSM, Subset 260 & {\color{red}3.8356} & 134.5 \\
		\hline
	\end{tabular}\caption{The RSTE and CPU time associated with the data subset model (DSM) based predictions using Algorithm 1 using the $N = 59,488$ training observations $14,873$ validation observations from \citet{bradley2014_comp}. Also, the RSTE and CPU time associated with fixed rank kriging, lattice kriging, negative exponential distance weighting (EDW), and the stochastic partial differential equation approach from \citet{lindgren-2011}. Small values are highlighted in red. }\label{tab:1}
\end{center}
\end{table}

\subsection{Application: Moderate Resolution Imaging Spectroradiometer Cloud Data} We consider a high-dimensional Bernoulli dataset analyzed in \citet{bradleyLCM}. This Bernoulli dataset consists of data obtained by the Moderate Resolution Imaging Spectroradiometer (MODIS) on December 18, 1999. MODIS is a remote sensing instrument that is on-board the Terra satellite, which is also a part of NASA's EOS. The raw data consists of measures of spectral radiances defined on a 1 km $\times$ 1 km grid computed using cloud detection algorithms \citep{aritrajsm}. These measurements are then thresholded to be either zero or one to indicate whether or not a cloud is present. This dataset is extremely large with $N = 2, 748, 620$.


The covariates and bisquare radial basis functions from \citet{aritrajsm} and \citet{bradleyLCM} are used for illustration. The sun-glint, intercept and 127 bisquare basis functions are used as covariates. We let $G=10,000$ MCMC iterations with a burn-in of $g_{0}=1000$. Trace plots are informally checked, with no lack of convergence detected. Simple random sampling without replacement is used to define $Pr(\bm{\delta}\vert n)$.

Five percent of the observations are considered hold-out observations. As a naive classifier, the predicted values are thresholded based on the midpoint of the range of values of the predicted probability of clear skies. False positives and false negatives over the  $5\%$ hold-out observations are used to validate the models. In Figure 4, we plot the data, and the predicted values. Under each panel we give the corresponding misclassification rates and CPU time. Here, we that for all $n$, the predicted probabilities reflect the pattern in the data. Additionally, setting $n = N$ leads to little gains in misclassification rates, no obvious change in the patterns of the predictions, and noticeably longer CPU times. Thus, our approach appears to have comparable out-of-sample and in-sample performance, and has a clear computational advantage. 

\begin{figure}
	\begin{center}
	\begin{tabular}{c}
	\includegraphics[height=2.5in, width=6in]{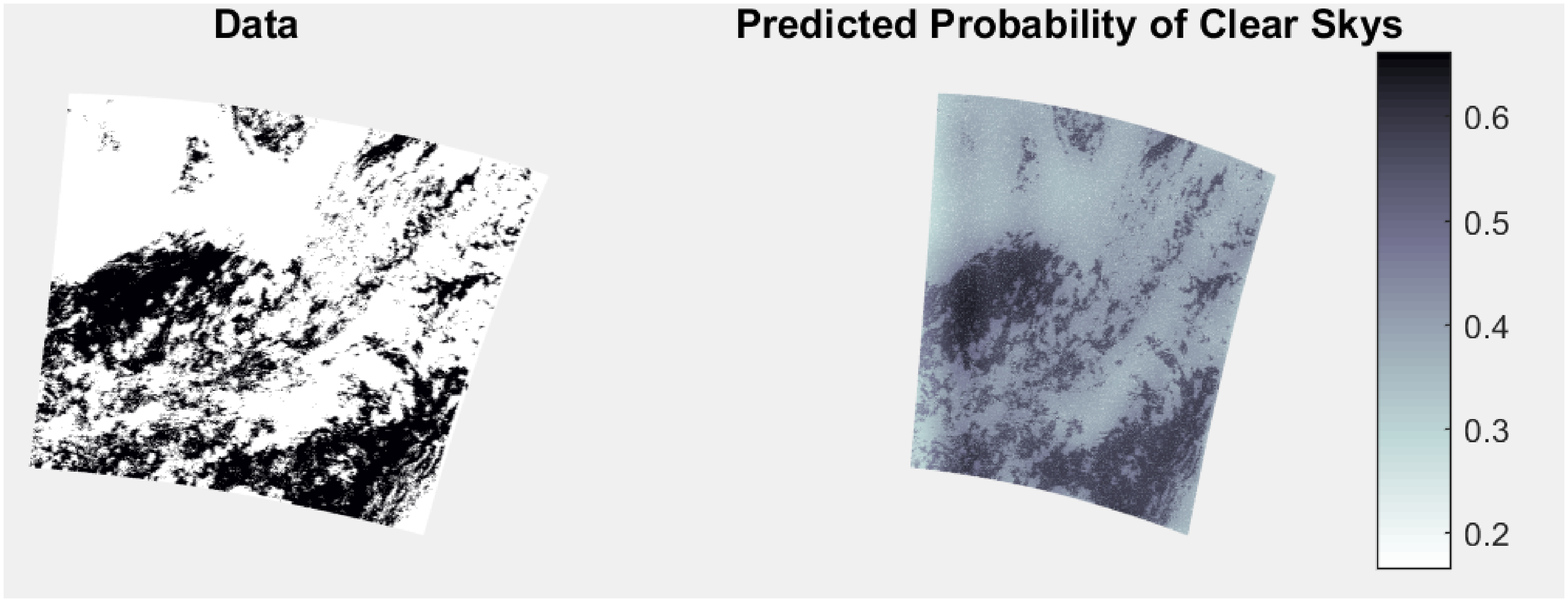}
\end{tabular}
\end{center}

	\begin{tabular}{ccc}
		\includegraphics[height=2.5in, width=2in]{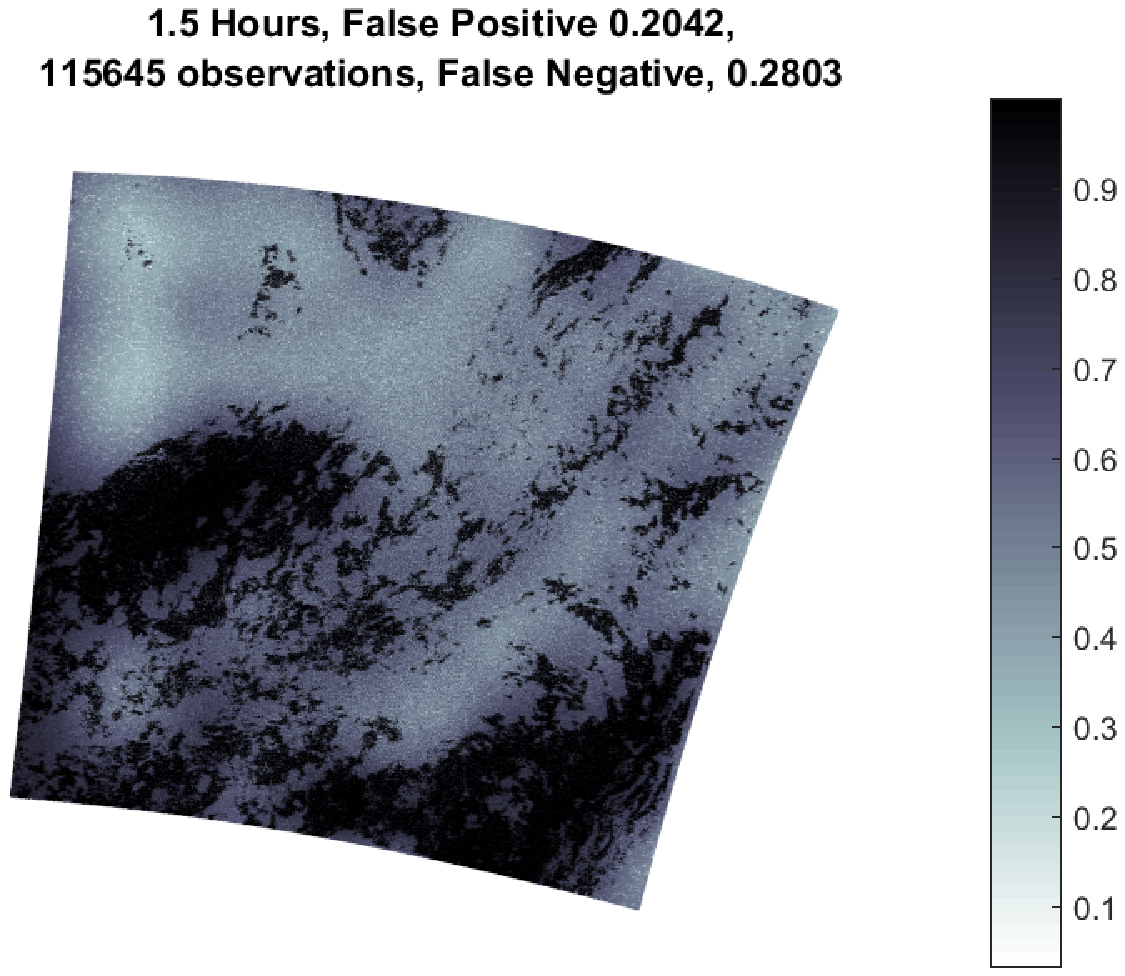}&
		\includegraphics[height=2.5in, width=2in]{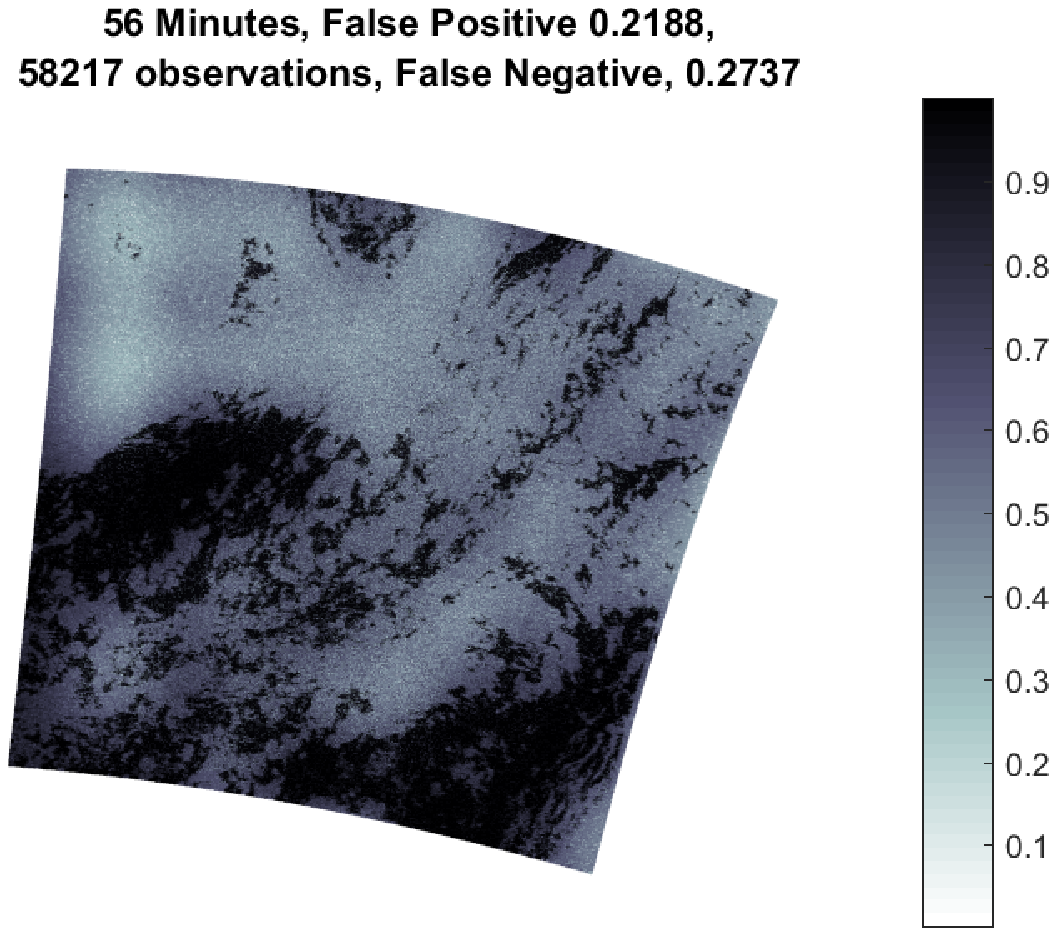}&
		\includegraphics[height=2.5in, width=2in]{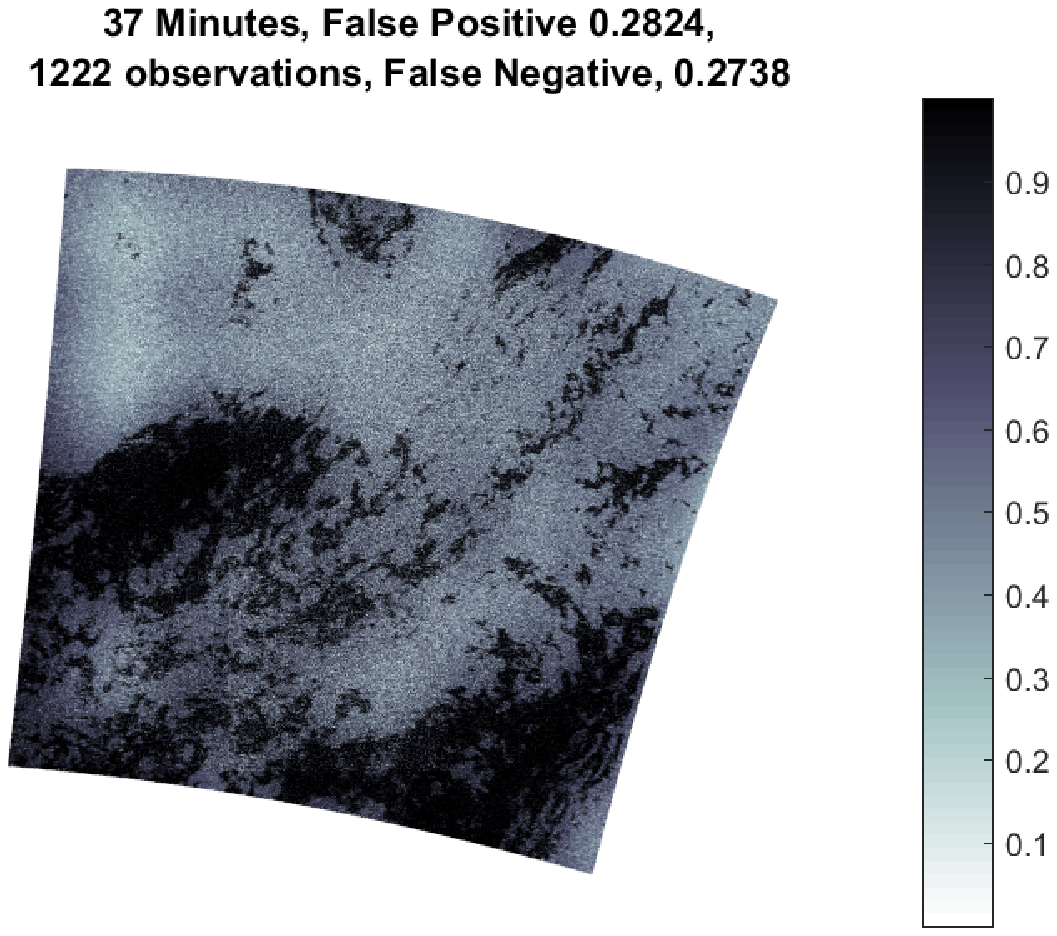}
	\end{tabular}
	\caption{The top two panels display the observed data, and the predictions from \citet{bradleyLCM} using all $2.75$ million observations. The predictions in the top right panel took 12 hours. The false positive rate and false negative rates for these predictions are 0.2774 and 0.2163. The bottom panels give predictions based on three different choices of $n$. The title headings of the bottom three panels includes the associated CPU times, false positive rates, and false negative rates.}\label{fig:4}
\end{figure}
%

\section{Discussion} In this article, we propose a class of Bayesian hierarchical models that includes a hyperparameter that allows one to answer the question posed in the title of this article. Specifically, we consider subsampling the data, and we incorporate this subjectivity directly into the Bayesian model through a subset model and a modified data model, which we call the data subset model. Specifically, the subset model introduces Bernoulli random variables that is used to remove or keep a datum's contribution to the data model. This data subset model is specified in a way so that the marginal distribution of the dataset is unaffected by subsampling, and so that the Bernoulli random variables are independent of the observed data. Thus, we interpret these Bernoulli random variables as modeler induced error terms. We show that our specification implicitly adds the assumption that a subsample of the dataset is partially sufficient for the unknown parameters in the expression of the likelihood. Furthermore, posterior inference from the reduced parametric model does not place more restrictive assumptions on the full parametric model, but rather, \textit{removes} assumptions on the full parametric model. 

In our simulation study we found that one sees little gains in prediction errors as $n$ increases. Additionally values of $n$ specified ``too small'' tend to produce over smoothed and noisy estimates. Two applications showed that dramatic increases in computation time can be achieved with little to no consequences on the in-sample and out-sample performances.

There several important concepts to keep in mind when implementing this methodology. First, the best predictor that you can compute in 5 minutes (or some other prespecified time) may not perform well in practice. This may possibly be because $n$ observations may not be partially sufficient (see Proposition 2). Furthermore, there is no guarantee that your time constraint is possible. That is, one has to specify a sequence of values for $n$, and if the first value in the sequence takes longer than five minutes than your goal can not be achieved using this method. Similarly, it takes longer than five minutes to find the best predictor that can be computed in five minutes. Specifically, the last value in the sequence of $n$ will take the longest, and may take longer than desired. Finally, the answer to the question in the title changes with the computer. This shifts some of the computational burden from the statistical model to the computer.

\section*{Appendix} 

\noindent
\textit{Proof of Proposition 1:} The joint distribution associated with (\ref{subhmnotation}) starts by multiplying the parameter model, process model, data subset model, and subset model to obtain:
\begin{align*}Pr(\bm{\delta}\vert n){ \left\lbrace\prod_{i = 1}^{N} f(Y_{i}\vert \bm{\nu},\bm{\theta},\bm{\delta})^{\delta_{i}}\right\rbrace f(\bm{\nu}\vert \bm{\theta})f(\bm{\theta})}\hspace{2pt} \frac{m(\bm{1}_{N},\textbf{y})}{{m(\bm{\delta},\textbf{y})}}.
\end{align*}
\noindent
To obtain the marginal distribution, integrate across $\bm{\nu}$ and $\bm{\theta}$, and sum across $\bm{\delta}$ to obtain,
\begin{align*}\sum_{\bm{\delta}\in \Delta}Pr(\bm{\delta}\vert n){\int \int \left\lbrace\prod_{i = 1}^{N} f(Y_{i}\vert \bm{\nu},\bm{\theta},\bm{\delta})^{\delta_{i}}\right\rbrace f(\bm{\nu}\vert \bm{\theta})f(\bm{\theta})d\bm{\nu}d\bm{\theta}}\hspace{2pt} \frac{m(\bm{1}_{N},\textbf{y})}{{m(\bm{\delta},\textbf{y})}} = m(\bm{1}_{N},\textbf{y}).
\end{align*}
\noindent
Since $m(\bm{1}_{N},\textbf{y})$ is assumed proper, we have that the model in (\ref{hmnotation}) is proper. This completes the result.\\

\noindent
\textit{Proof of Proposition 2:} From (\ref{subhmnotation}), the conditional distribution

\begin{equation*}
f(\textbf{y}\vert \textbf{y}_{\delta},\bm{\nu},\bm{\theta},\bm{\delta}) = \frac{\left\lbrace\prod_{i = 1}^{N} f(Y_{i}\vert \bm{\nu},\bm{\theta},\bm{\delta})^{\delta_{i}} \right\rbrace\hspace{5pt} \frac{m\left(\bm{1}_{N},\textbf{y}\right)}{m(\bm{\delta}, \textbf{y}_{\delta})}}{\int \left\lbrace\prod_{i = 1}^{N} f(Y_{i}\vert \bm{\nu},\bm{\theta},\bm{\delta})^{\delta_{i}} \right\rbrace\hspace{5pt} \frac{m\left(\bm{1}_{N},\textbf{y}\right)}{m(\bm{\delta}, \textbf{y}_{\delta})}d\textbf{y}_{-\delta}} = \frac{m(\bm{1}_{N},\textbf{y})}{\int m(\bm{1}_{N},\textbf{y}) d\textbf{y}_{-\delta}}
\end{equation*}
\noindent
Thus,
\begin{equation*}
f(\textbf{y}\vert \bm{\nu},\bm{\theta},\bm{\delta}) = f(\textbf{y}_{\delta}\vert \bm{\nu},\bm{\theta},\bm{\delta})\frac{m(\bm{1}_{N},\textbf{y})}{\int m(\bm{1}_{N},\textbf{y}) d\textbf{y}_{-\delta}},
\end{equation*}
\noindent
which satisfies (\ref{partial}) with $T(\textbf{y}) = \textbf{y}_{\delta}$, $h(T(\textbf{y}), \bm{\nu},\bm{\theta}, \bm{\delta}) = f(\textbf{y}_{\delta}\vert \bm{\nu},\bm{\theta},\bm{\delta})$ and $g(\textbf{y},\bm{\delta}) = \frac{m(\bm{1}_{N},\textbf{y})}{\int m(\bm{1}_{N},\textbf{y}) d\textbf{y}_{-\delta}}$.\\

\noindent
\textit{Proof of Proposition 3:} The joint distribution associated with (\ref{subhmnotation}) starts by multiplying the parameter model, process model, data subset model, and subset model to obtain:
\begin{align*}Pr(\bm{\delta}\vert n){ \left\lbrace\prod_{i = 1}^{N} f(Y_{i}\vert \bm{\nu},\bm{\theta},\bm{\delta})^{\delta_{i}}\right\rbrace f(\bm{\nu}\vert \bm{\theta})f(\bm{\theta})}\hspace{2pt} \frac{m(\bm{1}_{N},\textbf{y})}{{m(\bm{\delta},\textbf{y})}}.
\end{align*}
\noindent
Then to obtain the joint distribution of $\bm{\delta}$ and $\textbf{y}$, integrate across $\bm{\nu}$ and $\bm{\theta}$ to obtain,
\begin{align*}Pr(\bm{\delta}\vert n){\int \int \left\lbrace\prod_{i = 1}^{N} f(Y_{i}\vert \bm{\nu},\bm{\theta},\bm{\delta})^{\delta_{i}}\right\rbrace f(\bm{\nu}\vert \bm{\theta})f(\bm{\theta})d\bm{\nu}d\bm{\theta}}\hspace{2pt} \frac{m(\bm{1}_{N},\textbf{y})}{{m(\bm{\delta},\textbf{y})}} = Pr(\bm{\delta}\vert n)m(\bm{1}_{N},\textbf{y}),
\end{align*}
\noindent
which completes the result.\\

\noindent
\textit{Proof of Proposition 4:} From (\ref{sphmnotation}), the conditional distribution

\begin{equation*}
f(\textbf{y}\vert \textbf{y}_{\delta},\bm{\nu},\bm{\theta},\bm{\delta}) = \frac{\left\lbrace\prod_{i = 1}^{N} f(Y_{i}\vert \bm{\nu},\bm{\theta},\bm{\delta})^{\delta_{i}} \right\rbrace\hspace{5pt} w_{\delta}(\textbf{y}_{-\delta})}{\int \left\lbrace\prod_{i = 1}^{N} f(Y_{i}\vert \bm{\nu},\bm{\theta},\bm{\delta})^{\delta_{i}} \right\rbrace\hspace{5pt} w_{\delta}(\textbf{y}_{-\delta})d\textbf{y}_{-\delta}} = \frac{w_{\delta}(\textbf{y}_{-\delta})}{\int w_{\delta}(\textbf{y}_{-\delta}) d\textbf{y}_{-\delta}}
\end{equation*}
\noindent
Thus,
\begin{equation*}
f(\textbf{y}\vert \bm{\nu},\bm{\theta},\bm{\delta}) = f(\textbf{y}_{\delta}\vert \bm{\nu},\bm{\theta},\bm{\delta})\frac{w_{\delta}(\textbf{y}_{-\delta})}{\int w_{\delta}(\textbf{y}_{-\delta}) d\textbf{y}_{-\delta}},
\end{equation*}
\noindent
which satisfies (\ref{partial}) with $T(\textbf{y}) = \textbf{y}_{\delta}$, $h(T(\textbf{y}), \bm{\nu},\bm{\theta}, \bm{\delta}) = f(\textbf{y}_{\delta}\vert \bm{\nu},\bm{\theta},\bm{\delta})$ and $g(\textbf{y},\bm{\delta}) = \frac{w_{\delta}(\textbf{y}_{-\delta})}{\int w_{\delta}(\textbf{y}_{-\delta}) d\textbf{y}_{-\delta}} = w_{\delta}(\textbf{y}_{-\delta})$.\\

\noindent
\textit{Proof of Proposition 5:}
We have,
\begin{align*}
f_{SFM}(\bm{\nu},\bm{\theta}\vert \textbf{y},\bm{\delta},n)
&= \frac{ \left\lbrace\prod_{i = 1}^{N} f(Y_{i}\vert \bm{\nu},\bm{\theta},\bm{\delta})^{\delta_{i}}\right\rbrace w_{\delta}(\textbf{y}_{-\delta})f(\bm{\nu}\vert \bm{\theta}) f(\bm{\theta})}{w_{\delta}(\textbf{y}_{-\delta})m(\bm{\delta},\textbf{y})} = \frac{ \left\lbrace\prod_{i = 1}^{N} f(Y_{i}\vert \bm{\nu},\bm{\theta},\bm{\delta})^{\delta_{i}}\right\rbrace f(\bm{\nu}\vert \bm{\theta}) f(\bm{\theta}) }{m(\bm{\delta},\textbf{y})}\\
& = f_{DSM}(\bm{\nu},\bm{\theta}\vert \textbf{y},\bm{\delta},n),
\end{align*}
\noindent
which completes the proof.\\

\noindent
\textit{Proof of Proposition 6:}
We have,
\begin{align*}
f_{SFM}(\bm{\nu},\bm{\theta}\vert \textbf{y},n)&=\underset{\bm{\delta}\in \Delta}{\sum} Pr(\bm{\delta}\vert \textbf{y}, n)f_{SFM}(\bm{\nu},\bm{\theta}\vert \textbf{y},\bm{\delta},n)\\
&=\underset{\bm{\delta}\in \Delta}{\sum} Pr(\bm{\delta}\vert n)f_{SFM}(\bm{\nu},\bm{\theta}\vert \textbf{y},\bm{\delta},n)\\
&=\underset{\bm{\delta}\in \Delta}{\sum} Pr(\bm{\delta}\vert n)f_{DSM}(\bm{\nu},\bm{\theta}\vert \textbf{y},\bm{\delta},n)\\
& = f_{DSM}(\bm{\nu},\bm{\theta}\vert \textbf{y},n),
\end{align*}
\noindent
where the second equality holds by the assumption of independence between $\textbf{y}$ and $\bm{\delta}$, the third equality holds by Proposition 4, and the last equality holds by Proposition 3.

\section*{Acknowledgments} This research was partially supported by the U.S. National Science Foundation (NSF) under NSF grant SES-1853099.
  
  \singlespacing
\bibliographystyle{jasa} 
\bibliography{myref33}

\begin{thebibliography}{69}
\newcommand{\enquote}[1]{``#1''}
\expandafter\ifx\csname natexlab\endcsname\relax\def\natexlab#1{#1}\fi

\bibitem[\protect\citename{Anderson and Dean, }2014]{ANDERSONsub}
Anderson, L.~C. and Dean, N. (2014).
\newblock \enquote{Identifying clusters in Bayesian disease mapping.}
\newblock {\em Biostatistics\/}, 15, 457--469.

\bibitem[\protect\citename{Banerjee et~al., }2015]{banerjee-etal-2004}
Banerjee, S., Carlin, B.~P., and Gelfand, A.~E. (2015).
\newblock {\em Hierarchical Modeling and Analysis for Spatial Data\/}.
\newblock London, UK: Chapman and Hall.

\bibitem[\protect\citename{Banerjee et~al., }2008]{banerjee}
Banerjee, S., Gelfand, A.~E., Finley, A.~O., and Sang, H. (2008).
\newblock \enquote{Gaussian predictive process models for large spatial data
  sets.}
\newblock {\em Journal of the Royal Statistical Society, Series B\/}, 70,
  825--848.

\bibitem[\protect\citename{Barbian and Assuncao, }2017]{BARBIAN}
Barbian, M.~H. and Assuncao, R.~M. (2017).
\newblock \enquote{Spatial subsemble estimator for large geostatistical data.}
\newblock {\em Spatial Statistics\/}, 22, 68--88.

\bibitem[\protect\citename{Bardenet et~al., }2014]{Bardenet}
Bardenet, R., Doucet, A., and Holmes, C.~H. (2014).
\newblock \enquote{Towards scaling up Markov chain Monte Carlo: An adaptive
  subsampling approach.}
\newblock In {\em In Proceedings of the 30th International Conference on
  Machine Learning (ICML'14)\/},  405--413.

\bibitem[\protect\citename{Barry, }1986]{barry1986nonparametric}
Barry, D. (1986).
\newblock \enquote{Nonparametric Bayesian regression.}
\newblock {\em The Annals of Statistics\/}, 14, 3, 934--953.

\bibitem[\protect\citename{Basu, }1978]{BasuReview}
Basu, D. (1978).
\newblock \enquote{On Partial Sufficiency: A Review.}
\newblock {\em The Journal of Statistical Planning and Inference\/}, 2, 1--13.

\bibitem[\protect\citename{Berliner, }1996]{berlinhier}
Berliner, L.~M. (1996).
\newblock {\em Hierarchical Bayesian Time-Series Models\/}.
\newblock Kluwer Academic Publishers, Dordrecht, NL.

\bibitem[\protect\citename{Besag et~al., }1991]{besag-91}
Besag, J., York, J., and Molli\'e, A. (1991).
\newblock \enquote{Bayesian image restoration, with two applications in spatial
  statistics.}
\newblock {\em Annals of the Institute of Statistical Mathematics\/}, 43,
  1--20.

\bibitem[\protect\citename{Besag, }1974]{besag-74}
Besag, J.~E. (1974).
\newblock \enquote{{Spatial interaction and the statistical analysis of lattice
  systems (with discussion)}.}
\newblock {\em Journal of the Royal Statistical Society, Series B\/}, 36,
  192--236.

\bibitem[\protect\citename{Besag, }1986]{besag-86}
--- (1986).
\newblock \enquote{{On the statistical analysis of dirty pictures (with
  discussion)}.}
\newblock {\em Journal of the Royal Statistical Society, Series B\/}, 48,
  259--302.

\bibitem[\protect\citename{Bien and Tibshirani, }2011]{bien2011sparse}
Bien, J. and Tibshirani, R.~J. (2011).
\newblock \enquote{Sparse estimation of a covariance matrix.}
\newblock {\em Biometrika\/}, 98, 4, 807--820.

\bibitem[\protect\citename{Bradley et~al., }2015{\natexlab{a}}]{bradleyMSTM}
Bradley, J., Holan, S., and Wikle, C. (2015{\natexlab{a}}).
\newblock \enquote{Multivariate spatio-temporal models for high-dimensional
  areal data with application to Longitudinal Employer-Household Dynamics.}
\newblock {\em The Annals of Applied Statistics\/}, 9, 1761--1791.

\bibitem[\protect\citename{Bradley et~al., }2011]{bradley2011}
Bradley, J.~R., Cressie, N., and Shi, T. (2011).
\newblock \enquote{Selection of rank and basis functions in the Spatial Random
  Effects model.}
\newblock In {\em Proceedings of the 2011 Joint Statistical Meetings\/},
  3393--3406. Alexandria, VA: American Statistical Association.

\bibitem[\protect\citename{Bradley et~al., }2015{\natexlab{b}}]{bradleyTEST}
--- (2015{\natexlab{b}}).
\newblock \enquote{Comparing and selecting spatial predictors using local
  criteria.}
\newblock {\em TEST\/}, 24, 1--28.

\bibitem[\protect\citename{Bradley et~al., }2016]{bradley2014_comp}
--- (2016).
\newblock \enquote{A comparison of spatial predictors when datasets could be
  very large.}
\newblock {\em Statistics Surveys\/}, 10, 100--131.

\bibitem[\protect\citename{Bradley et~al., }2019{\natexlab{a}}]{bradleyLCM}
Bradley, J.~R., Holan, S.~H., and Wikle, C.~K. (2019{\natexlab{a}}).
\newblock \enquote{Bayesian Hierarchical Models with Conjugate Full-Conditional
  Distributions for Dependent Data from the Natural Exponential Family.}
\newblock {\em arXiv e-prints: 1701.07506\/}.

\bibitem[\protect\citename{Bradley et~al., }2019{\natexlab{b}}]{bradleyJTSA}
Bradley, J.~R., Wikle, C.~K., and Holan, S.~H. (2019{\natexlab{b}}).
\newblock \enquote{Spatio-Temporal Models for Big Multinomial Data using the
  Conditional Multivariate Logit-Beta Distribution.}
\newblock {\em Journal of Time Series Analysis\/}.

\bibitem[\protect\citename{Broderick et~al., }2013]{broderick2013streaming}
Broderick, T., Boyd, N., Wibisono, A., Wilson, A.~C., and Jordan, M.~I. (2013).
\newblock \enquote{Streaming variational bayes.}
\newblock In {\em Advances in Neural Information Processing Systems\/},
  1727--1735.

\bibitem[\protect\citename{Cressie and Johannesson, }2006]{johan-2006}
Cressie, N. and Johannesson, G. (2006).
\newblock \enquote{Spatial prediction for massive data sets.}
\newblock In {\em Australian Academy of Science Elizabeth and Frederick White
  Conference\/},  1--11. Australian Academy of Science, Canberra.

\bibitem[\protect\citename{Cressie and Johannesson, }2008]{johan}
--- (2008).
\newblock \enquote{Fixed rank kriging for very large spatial data sets.}
\newblock {\em Journal of the Royal Statistical Society, Series B\/}, 70,
  209--226.

\bibitem[\protect\citename{Cressie et~al.,
  }2010{\natexlab{a}}]{cressie-shi-kang-2010}
Cressie, N., Shi, T., and Kang, E.~L. (2010{\natexlab{a}}).
\newblock \enquote{{ Fixed Rank Filtering for spatio-temporal data}.}
\newblock {\em Journal of Computational and Graphical Statistics\/}, 19,
  724--745.

\bibitem[\protect\citename{Cressie et~al.,
  }2010{\natexlab{b}}]{kang-cressie-shi-2010}
--- (2010{\natexlab{b}}).
\newblock \enquote{{ Using temporal variability to improve spatial mapping with
  application to satellite data}.}
\newblock {\em Canadian Journal of Statistics\/}, 38, 271--289.

\bibitem[\protect\citename{Cressie and Wikle, }2011]{cressie-wikle-book}
Cressie, N. and Wikle, C.~K. (2011).
\newblock {\em Statistics for Spatio-Temporal Data\/}.
\newblock Hoboken, NJ: Wiley.

\bibitem[\protect\citename{Dang et~al., }2019]{dang2019hamiltonian}
Dang, K.-D., Quiroz, M., Kohn, R., Tran, M.-N., and Villani, M. (2019).
\newblock \enquote{Hamiltonian Monte Carlo with energy conserving subsampling.}
\newblock {\em Journal of machine learning research\/}, 20, 100, 1--31.

\bibitem[\protect\citename{Finley et~al., }2009]{finley}
Finley, A.~O., Sang, H., Banerjee, S., and Gelfand, A.~E. (2009).
\newblock \enquote{Improving the performance of predictive process modeling for
  large datasets.}
\newblock {\em Computational Statistics and Data Analysis\/}, 53, 2873--2884.

\bibitem[\protect\citename{Friedman et~al., }2008]{friedman2008sparse}
Friedman, J., Hastie, T., and Tibshirani, R. (2008).
\newblock \enquote{Sparse inverse covariance estimation with the graphical
  lasso.}
\newblock {\em Biostatistics\/}, 9, 3, 432--441.

\bibitem[\protect\citename{Furrer et~al., }2006]{furrer2006covariance}
Furrer, R., Genton, M.~G., and Nychka, D. (2006).
\newblock \enquote{Covariance tapering for interpolation of large spatial
  datasets.}
\newblock {\em Journal of Computational and Graphical Statistics\/}, 15, 3,
  502--523.

\bibitem[\protect\citename{Gao and Bradley, }2019]{gao2019bayesian}
Gao, H. and Bradley, J.~R. (2019).
\newblock \enquote{Bayesian analysis of areal data with unknown adjacencies
  using the stochastic edge mixed effects model.}
\newblock {\em Spatial Statistics\/}.

\bibitem[\protect\citename{Gelfand and Smith, }1990]{gelfand1990sampling}
Gelfand, A.~E. and Smith, A.~F. (1990).
\newblock \enquote{Sampling-based approaches to calculating marginal
  densities.}
\newblock {\em Journal of the American statistical association\/}, 85, 410,
  398--409.

\bibitem[\protect\citename{Geman and Geman, }1984]{Geman1984stochastic}
Geman, S. and Geman, D. (1984).
\newblock \enquote{Stochastic relaxation, Gibbs distributions, and the Bayesian
  restoration of images.}
\newblock {\em IEEE Transactions on pattern analysis and machine
  intelligence\/}, , 6, 721--741.

\bibitem[\protect\citename{Guinness, }2018]{guinness2018permutation}
Guinness, J. (2018).
\newblock \enquote{Permutation and grouping methods for sharpening Gaussian
  process approximations.}
\newblock {\em Technometrics\/}, 60, 4, 415--429.

\bibitem[\protect\citename{Gunawan et~al., }2017]{gunawan2017fast}
Gunawan, D., Tran, M.-N., and Kohn, R. (2017).
\newblock \enquote{Fast inference for intractable likelihood problems using
  variational Bayes.}
\newblock {\em arXiv preprint arXiv:1705.06679\/}.

\bibitem[\protect\citename{Heaton et~al., }2017]{HEATONsub}
Heaton, M.~J., Christensen, W.~F., and Terres, M.~A. (2017).
\newblock \enquote{Nonstationary Gaussian process models using spatial
  hierarchical clustering from finite differences.}
\newblock {\em Technometrics\/}, 59, 93--101.

\bibitem[\protect\citename{Heaton et~al., }2018]{heaton2018case}
Heaton, M.~J., Datta, A., Finley, A.~O., Furrer, R., Guinness, J., Guhaniyogi,
  R., Gerber, F., Gramacy, R.~B., Hammerling, D., Katzfuss, M., and Lindgren,
  F. (2018).
\newblock \enquote{A case study competition among methods for analyzing large
  spatial data.}
\newblock {\em Journal of Agricultural, Biological and Environmental
  Statistics\/},  1--28.

\bibitem[\protect\citename{Huggins et~al., }2016]{huggins2016coresets}
Huggins, J., Campbell, T., and Broderick, T. (2016).
\newblock \enquote{Coresets for scalable Bayesian logistic regression.}
\newblock In {\em Advances in Neural Information Processing Systems\/},
  4080--4088.

\bibitem[\protect\citename{Jolliffe, }1973]{jolliffe1973discarding}
Jolliffe, I.~T. (1973).
\newblock \enquote{Discarding variables in a principal component analysis. II:
  Real data.}
\newblock {\em Journal of the Royal Statistical Society: Series C (Applied
  Statistics)\/}, 22, 1, 21--31.

\bibitem[\protect\citename{Kang and Cressie, }2011]{kang-cressie-2011}
Kang, E.~L. and Cressie, N. (2011).
\newblock \enquote{{Bayesian inference for the spatial random effects model}.}
\newblock {\em Journal of the American Statistical Association\/}, 106, 972 --
  983.

\bibitem[\protect\citename{Katzfuss and Cressie, }2011]{katzfuss_1}
Katzfuss, M. and Cressie, N. (2011).
\newblock \enquote{{Spatio-temporal smoothing and EM estimation for massive
  remote-sensing data sets}.}
\newblock {\em Journal of Time Series Analysis\/}, 32, 430--446.

\bibitem[\protect\citename{Katzfuss and Cressie, }2012]{katzfuss2012}
--- (2012).
\newblock \enquote{Bayesian hierarchical spatio-temporal smoothing for very
  large datasets.}
\newblock {\em Environmetrics\/}, 23, 94--107.

\bibitem[\protect\citename{Katzfuss and Guinness, }2017]{katzfuss2017general}
Katzfuss, M. and Guinness, J. (2017).
\newblock \enquote{A general framework for Vecchia approximations of Gaussian
  processes.}
\newblock {\em arXiv preprint arXiv:1708.06302\/}.

\bibitem[\protect\citename{Kim et~al., }2005]{kim2}
Kim, H.~M., Mallick, B.~K., and Holmes, C. (2005).
\newblock \enquote{Analyzing nonstationary spatial data using piecewise
  Gaussian processes.}
\newblock {\em Journal of the American Statistical Association\/}, 100, 653 --
  668.

\bibitem[\protect\citename{Kleiner et~al., }2012]{kleiner2012big}
Kleiner, A., Talwalkar, A., Sarkar, P., and Jordan, M.~I. (2012).
\newblock \enquote{The big data bootstrap.}
\newblock {\em arXiv preprint arXiv:1206.6415\/}.

\bibitem[\protect\citename{Knorr-Held and Rasser, }2000]{KNORR}
Knorr-Held, L. and Rasser, G. (2000).
\newblock \enquote{Bayesian detection of clusters and discontinuities in
  disease maps.}
\newblock {\em Biometrics\/}, 56, 13 -- 21.

\bibitem[\protect\citename{Konomi et~al., }2014]{KONOMI}
Konomi, B.~A., Sang, H., and Mallick, B.~K. (2014).
\newblock \enquote{Adaptive Bayesian nonstationary modeling for large spatial
  datasets using covariance approximations.}
\newblock {\em Journal of Computational and Graphical Statistics\/}, 23, 802 --
  829.

\bibitem[\protect\citename{Korattikara et~al., }2014]{korattikara2014austerity}
Korattikara, A., Chen, Y., and Welling, M. (2014).
\newblock \enquote{Austerity in MCMC land: Cutting the Metropolis-Hastings
  budget.}
\newblock In {\em International Conference on Machine Learning\/},  181--189.

\bibitem[\protect\citename{Liang et~al., }2013]{LIANGsub}
Liang, F., Cheng, Y., Song, Q., Park, J., and Yang, P. (2013).
\newblock \enquote{A resampling based stochastic approximation method for
  analysis of large geostatistical data.}
\newblock {\em Journal of the American Statistical Association\/}, 108,
  325--339.

\bibitem[\protect\citename{Lindgren et~al., }2011]{lindgren-2011}
Lindgren, F., Rue, H., and Lindstr\"{o}m, J. (2011).
\newblock \enquote{An explicit link between Gaussian fields and Gaussian Markov
  random fields: The stochastic partial differential equation approach.}
\newblock {\em Journal of the Royal Statistical Society, Series B\/}, 73,
  423--498.

\bibitem[\protect\citename{Lohr, }1999]{lohr-survey}
Lohr, S. (1999).
\newblock {\em Sampling Design and Analysis\/}.
\newblock Pacific Grove, CA, USA: Brooks/Cole Publishing Company.

\bibitem[\protect\citename{Lohr and Brick, }2012]{lohr}
Lohr, S. and Brick, M. (2012).
\newblock \enquote{Blending domain estimates from two victimization surveys
  with possible bias.}
\newblock {\em The Canadian Journal of Statistics\/}, 40, 679--969.

\bibitem[\protect\citename{Maclaurin and Adams, }2014]{Firefly2014}
Maclaurin, D. and Adams, R.~P. (2014).
\newblock \enquote{Firefly Monte Carlo: Exact MCMC with Subsets of Data.}
\newblock {\em arXiv: 1403.5693\/}.

\bibitem[\protect\citename{Neelon et~al., }2014]{NEELON}
Neelon, B., Gelfand, A.~E., and Miranda, M.~L. (2014).
\newblock \enquote{A multivariate spatial mixture model for areal data:
  examining regional differences in standardized test scores.}
\newblock {\em Journal of the Royal Statistical Society: Series C\/}, 63, 737
  -- 761.

\bibitem[\protect\citename{Nychka et~al., }2015]{nychkaLK}
Nychka, D., Bandyopadhyay, S., Hammerling, D., Lindgren, F., and Sain, S.
  (2015).
\newblock \enquote{A multi-resolution Gaussian process model for the analysis
  of large spatial data sets.}
\newblock {\em Journal of Computational and Graphical Statistics\/},  DOI:
  10.1080/10618600.2014.914946.

\bibitem[\protect\citename{Quiroz et~al., }2019]{quiroz2019speeding}
Quiroz, M., Kohn, R., Villani, M., and Tran, M.-N. (2019).
\newblock \enquote{Speeding up MCMC by efficient data subsampling.}
\newblock {\em Journal of the American Statistical Association\/}, 114, 526,
  831--843.

\bibitem[\protect\citename{Reiss et~al., }2012]{reiss2012heterogeneity}
Reiss, C., Tumanov, A., Ganger, G.~R., Katz, R.~H., and Kozuch, M.~A. (2012).
\newblock \enquote{Heterogeneity and dynamicity of clouds at scale: Google
  trace analysis.}
\newblock In {\em Proceedings of the Third ACM Symposium on Cloud Computing\/},
  ~7. ACM.

\bibitem[\protect\citename{Rothman et~al., }2010]{rothman2010sparse}
Rothman, A.~J., Levina, E., and Zhu, J. (2010).
\newblock \enquote{Sparse multivariate regression with covariance estimation.}
\newblock {\em Journal of Computational and Graphical Statistics\/}, 19, 4,
  947--962.

\bibitem[\protect\citename{Rue and Held, }2005]{Rue2005gaussian}
Rue, H. and Held, L. (2005).
\newblock {\em Gaussian Markov random fields: theory and applications\/}.
\newblock CRC press.

\bibitem[\protect\citename{Sang and Huang, }2012]{sang}
Sang, H. and Huang, J. (2012).
\newblock \enquote{A full-scale approximation of covariance functions for large
  spatial data sets.}
\newblock {\em Journal of the Royal Statistical Society: Series B\/}, 74,
  111--132.

\bibitem[\protect\citename{Sengupta et~al., }2012]{aritrajsm}
Sengupta, A., Cressie, N., Frey, R., and Kahn, B. (2012).
\newblock \enquote{Statistical modeling of MODIS cloud data using the Spatial
  Random Effects model.}
\newblock In {\em Proceedings of the Joint Statistical Meetings\/},
  3111--3123. Alexandria, VA: American Statistical Association.

\bibitem[\protect\citename{Sethuraman, }1994]{sethuraman1994constructive}
Sethuraman, J. (1994).
\newblock \enquote{A constructive definition of Dirichlet priors.}
\newblock {\em Statistica Sinica\/},  639--650.

\bibitem[\protect\citename{Shi and Cressie, }2007]{cressie-tao}
Shi, T. and Cressie, N. (2007).
\newblock \enquote{{Global statistical analysis of MISR aerosol data: A massive
  data product from NASA's Terra satellite}.}
\newblock {\em Environmetrics\/}, 18, 665--680.

\bibitem[\protect\citename{Srivastava et~al., }2018]{srivastava2018scalable}
Srivastava, S., Li, C., and Dunson, D.~B. (2018).
\newblock \enquote{Scalable Bayes via barycenter in Wasserstein space.}
\newblock {\em The Journal of Machine Learning Research\/}, 19, 1, 312--346.

\bibitem[\protect\citename{Stein, }2014]{steinr}
Stein, M. (2014).
\newblock \enquote{Limitations on low rank approximations for covariance
  matrices of spatial data.}
\newblock {\em Spatial Statistics\/}, 8, 1--19.

\bibitem[\protect\citename{Stein et~al., }2004]{stein2004approximating}
Stein, M.~L., Chi, Z., and Welty, L.~J. (2004).
\newblock \enquote{Approximating likelihoods for large spatial data sets.}
\newblock {\em Journal of the Royal Statistical Society: Series B (Statistical
  Methodology)\/}, 66, 2, 275--296.

\bibitem[\protect\citename{Vecchia, }1988]{vecchia1988estimation}
Vecchia, A.~V. (1988).
\newblock \enquote{Estimation and model identification for continuous spatial
  processes.}
\newblock {\em Journal of the Royal Statistical Society: Series B
  (Methodological)\/}, 50, 2, 297--312.

\bibitem[\protect\citename{Wahba, }1990]{wahba}
Wahba, G. (1990).
\newblock {\em Spline Models for Observational Data\/}.
\newblock Philadelphia, PA: Society for Industrial and Applied Mathematics.

\bibitem[\protect\citename{Wikle, }2010]{wikleHandbook}
Wikle, C.~K. (2010).
\newblock \enquote{Low-rank representations for spatial processes.}
\newblock In {\em Handbook of Spatial Statistics\/}, eds. A.~E. Gelfand, P.~J.
  Diggle, M.~Fuentes, and P.~Guttorp,  107--118. Boca Raton, FL: Chapman $\&$
  Hall/CRC Press.

\bibitem[\protect\citename{Wikle and Cressie, }1999]{wiklecress_spt}
Wikle, C.~K. and Cressie, N. (1999).
\newblock \enquote{A dimension-reduced approach to space-time Kalman
  filtering.}
\newblock {\em Biometrika\/}, 86, 815--829.

\bibitem[\protect\citename{Zhao et~al., }2018]{zhao2018efficient}
Zhao, Y., Amemiya, Y., and Hung, Y. (2018).
\newblock \enquote{Efficient Gaussian process modeling using experimental
  design-based subagging.}
\newblock {\em Statistica Sinica\/}, 28, 3, 1459--1479.

\end{thebibliography}

\end{document}